\documentclass[usenatbib,useAMS]{mnras}
\usepackage{amsmath, amstext,amssymb}
\usepackage[flushleft]{threeparttable}
\usepackage{mathptmx, txfonts}
\usepackage[T1]{fontenc}
\usepackage{ae,aecompl}

\usepackage{bm}
\usepackage{graphicx}
\usepackage{multirow}
\usepackage{adjustbox}
\usepackage{textcomp}
\newcommand{\Ddt}{$D_{\Delta t}$}
\newcommand{\Dd}{$D_{{\rm d}}$}

\title[Kinematics and time delays]{Improving time-delay cosmography with spatially resolved kinematics}

\author[A. J. Shajib et al.]{
Anowar J. Shajib,$^{1}$\thanks{E-mail: ajshajib@astro.ucla.edu}
Tommaso Treu$^{1}$
and Adriano Agnello$^{2}$
\\
$^{1}$Department of Physics and Astronomy, University of California, Los Angeles, CA 90095-1547, USA\\
$^{2}$European Southern Observatory, Karl-Schwarzschild-Strasse 2, 85748 Garching bei M{\"u}nchen, Germany
}

\date{Accepted 2017 September 4. Received 2017 September 4; in original form 2017 February 14}

\pubyear{2017}

\begin{document}
\label{firstpage}
\pagerange{\pageref{firstpage}--\pageref{lastpage}}
\maketitle

\begin{abstract}
Strongly gravitational lensed quasars can be used to measure the
so-called time-delay distance \Ddt, and thus the Hubble constant $H_0$
and other cosmological parameters. Stellar kinematics of the deflector
galaxy play an essential role in this measurement by: (i) helping
break the mass-sheet degeneracy; (ii) determining in principle the
angular diameter distance \Dd\ to the deflector and thus further
improving the cosmological constraints. In this paper we simulate
observations of lensed quasars with integral field spectrographs and
show that spatially resolved kinematics of the deflector enable
further progress by helping break the mass-anisotropy
degeneracy. Furthermore, we use our simulations to obtain realistic
error estimates with current/upcoming instruments like OSIRIS on Keck
and NIRSPEC on the \textit{James Webb Space Telescope} for both distances
(typically $\sim6$ per cent on \Ddt\ and $\sim10$ per cent on \Dd). We use the error
estimates to compute cosmological forecasts for the sample of nine lenses
that currently have well measured time delays and deep \textit{Hubble Space
Telescope} images and for a sample of 40 lenses that is projected to be
available in a few years through follow-up of candidates found in
ongoing wide field surveys. We find that $H_0$ can be measured with
2 per cent (1 per cent) precision from nine (40) lenses in a flat $\Lambda$cold dark matter
cosmology. We study several other cosmological models beyond the flat
$\Lambda$cold dark matter model and find that time-delay lenses with spatially
resolved kinematics can greatly improve the precision of the
cosmological parameters measured by cosmic microwave background
data. 

\end{abstract}


\section{Introduction}

Our current understanding of cosmography, i.e. the description of
geometry and kinematics of the Universe, has been largely acquired
from the measurements of cosmic distances as a function of
redshift. For example, relative luminosity distance measurements using
Type Ia supernovae led to the discovery of dark energy \citep{Riess98,
  Perlmutter98}. More recently, baryon acoustic oscillation (BAO) in galaxy
clustering has been used to determine angular diameter distances as a
function of redshifts \citep{Eisenstein05, Shadab16}.

Absolute distances, and the Hubble constant $H_0$ in particular, play
a central role in cosmography. In
fact, the uncertainty on $H_0$ is currently one of the main limiting
factors in cosmological inferences based on the cosmic microwave
background \citep[CMB; e.g.][]{Suy++12, Wei++13}. The tension between
the recent measurement of the local value of $H_0$ to 2.4 per cent
precision determined from Type Ia supernovae \citep{Riess16, BVR16}, and that
extrapolated from the CMB assuming a flat $\Lambda$cold dark matter
($\Lambda$CDM) cosmology highlights the importance of absolute
distances. If the tension cannot be explained as residual systematic
uncertainties in one (or both) measurements, it may be an indication
of new physics, like additional families of relativistic particles,
departures from flatness, or dark energy that is not the cosmological
constant \citep{Riess16}.  In this context, independent and precise
measurements of absolute distances are needed to resolve this tension,
and may be required in order to disprove conclusively the standard
flat $\Lambda$CDM model.

Gravitational lens systems where the source is variable in time
provide a powerful direct measurement of distances, that is completely
independent of the local distance ladder and the CMB \citep{Refsdal64}. Substantial progress in data quality,
monitoring campaigns, and modelling techniques over the past decade has
finally allowed this technique to deliver on its promises
\citep[see][for a historical perspective and a review of current methods]{Treu16}. It has recently been shown that just three
lenses are sufficient to determine $H_0$ to 3.8 per cent precision
\citep[e.g.][]{Suyu10, Suy++13, Bonvin17}, in $\Lambda$CDM.

The primary distance measurement is the so-called time-delay distance
\Ddt, a multiplicative combination of the three angular diameter
distances between the observer, the deflector, and the source. By
combining the time-delay measurement with the stellar velocity
dispersion measurements of the deflector, it is possible to measure
also the angular diameter distance \Dd\ to the deflector
\citep{GLB08,Paraficz10, Jee15}, thereby improving the constraints on
the cosmological parameters \citep{Jee16}.

In order to harness the power of strong lenses to constrain
cosmography one needs to break two families of degeneracy. The first
one is the mass-sheet degeneracy \citep[MSD;][]{FGS85} and its generalizations
\citep{S+S13,SPT,XuEtal2016} that affect the interpretation of lensing
observables. Breaking this degeneracy requires making appropriate
physical assumptions on the mass profile of the main deflector
\citep{XuEtal2016} or on the properties of the source \citep{BAR16},
measuring the lensing effects along the line of sight
\citep{Suyu10,Suy++13,Gre++13,Slu++16,Rus++16}, and including as much
non-lensing information as possible, especially stellar velocity
dispersion of the deflector
\citep{T+K02b,Koo++03,Suyu10,Suy++13,Suy++14,Wong17}. The
interpretation of stellar velocity dispersion data introduces the
second degeneracy, known as the mass-anisotropy degeneracy \citep[see,
e.g.][and references therein]{vdM94,Cou++14}, whereby different combinations of mass profiles and stellar orbits can reproduce the same kinematic profiles. This holds especially for elliptical galaxies, which constitute most of the deflectors in strong lens samples. Even though lens galaxies and nearby ellipticals are on average consistent with simple density profiles and modest anisotropy \citep{Koo++09,Bar++11,AER14}, there are significant system-to-system variations and appreciable systematic uncertainties. Also the exploration of different anisotropy profiles can affect the inference on the mass profile, privileging regions of parameters space where the inferred masses depend weakly on the anisotropy parameters \citep[e.g. at large anisotropy radii,][]{Agn++14}, a problem that is exacerbated by kinematic data within the half-light radius. The mass-anisotropy degeneracy is alleviated in the \textit{virial regime} of large apertures \citep[e.g.][]{Treu02,AAE13}, so a combination of extended radial coverage and a tight control on systematics can be used to aid cosmography with lensing and stellar dynamics \citep[e.g.][]{BAR16}.

Spatially resolved kinematics of galaxy scale lensed quasars is
challenging with seeing limited observations, owing to the presence of
bright quasar within the typical separation of the order
of arcsecond. Diffraction limited spectroscopy is needed to make progress,
either from the ground with the assistance of laser guide star
adaptive optics (AO), or from space. Recent advances in AO
technology and the imminent launch of the \textit{James Webb Space Telescope} (\textit{JWST})
make this kind of measurement feasible, and calls for a detailed
investigation of its potential for cosmography.

In this paper, we investigate the improvements to time-delay
cosmography that can be expected in the next few years by combining
spatially resolved kinematics with lensing data. Unfortunately, state
of the art modelling techniques are too computationally expensive at
present to carry out a full-blown pixel-based analysis of a large
number of mock systems. Thus, in order to keep the computational cost
manageable, we develop a framework to simulate and model mock lenses
in a simplified and effective manner, but calibrated to yield
realistic uncertainties as they would be obtained with a pixel-based
analysis. We then apply these techniques to study the precision and
accuracy that can be achieved on \Dd\ and \Ddt\ per system for
plausible observational data quality that can be expected for current
\citep[e.g. OSIRIS on Keck,][]{Lar++06}, imminent (NIRSPEC on \textit{JWST}),
and future \citep[e.g. IRIS on the Thirty Metre Telescope
  (TMT),][]{Wri++16} integral field spectrographs (IFSs). Finally, we use our
results on the estimated precision of \Dd\ and \Ddt\ to forecast the
cosmological precision that can be attained with the current sample of
nine lenses for which accurate time delays and deep \textit{Hubble Space
Telescope} (\textit{HST}) imaging data are available, and for a sample of 40
lenses that is expected to be completed in the next few years by a
dedicated follow-up campaign of newly discovered lenses from ongoing
wide field imaging surveys [e.g. the STrong lensing Insights into the
Dark Energy Survey (STRIDES)\footnote{STRIDES is a Dark Energy Survey
  Broad External Collaboration; PI:
  Treu. \url{http://strides.astro.ucla.edu}}].

Our work builds upon and extends previous work by \citet{Jee16} in two
important ways. First, we consider spatially resolved kinematics
whereas \citet{Jee16} focused on integrated quantities. As we will
show, this aspect allows us to let anisotropy be a free parameter and
show that the mass-anisotropy degeneracy can be overcome. Secondly,
rather than assuming a target uncertainty on the two distances
\Ddt, \Dd\ [\citet{Jee16} adopted a fiducial 5 per cent uncertainty on both], we derive
them from realistic assumptions about the measurements exploring
different scenarios corresponding to variation in data quality, e.g.
effect of including kinematics, improved instrumental precision,
and observing conditions. We then use these uncertainties to infer
the attainable precisions on the cosmological parameters.

The structure of this paper is as follows. In \autoref{sect:model} we
briefly review the strong gravitational lensing formalism and describe
the mass models we used to simulate the deflector galaxy mass
distribution. In \autoref{sec:mock} we describe the methods to
create mock lensing and kinematic data from simulated strong lens
systems. We present our results on the precision of the cosmological
distances in \autoref{sect:distuncertainties} and forecast the
cosmological parameter uncertainties in \autoref{sect:cosmoinfer}. We
follow that with our discussion about the study and comparison with
previous works in \autoref{sect:discussion} and the limitations of this work in 
\autoref{sect:limitation}. Lastly, we conclude the
paper with a summary in \autoref{sect:summary}.

\section{Model ingredients}\label{sect:model}

Multiply-imaged quasars are ideal candidates for time-delay
cosmography as the time delay can be measured by monitoring quasar
variability. The deflector in such a system is usually an elliptical
galaxy. In this section, we first present a brief
review of the strong gravitational lensing formalism in \autoref{ssec:lens}.  Then in
\autoref{ssec:mass} we describe the models we use to simulate realistic
deflector mass distributions.

\subsection{Strong gravitational lensing}\label{ssec:lens}

In this subsection, we set the notation by briefly reviewing the
theory of strong gravitational lensing \cite[see][for a detailed description]{Schneider06}. Let us consider a strong gravitational-lens system with the deflector at the origin and the background source at $\bm{\beta}$. Then, the image positions $\bm{\theta}$ are given by the solutions of the lensing equation

\begin{equation}\label{eq:lenseq}
  \bm{\beta} = \bm{\theta} - \bm{\alpha}(\bm{\theta}),
\end{equation}
where $\bm{\alpha}(\bm{\theta}) = \bm{\nabla}_{\theta} \psi (\bm{\theta})$ is the deflection angle and $\psi$ is the deflection potential. The dimensionless quantity convergence $\kappa$ is defined as $\kappa(\bm{\theta}) \equiv \Sigma(\bm{\theta})/\Sigma_{{\rm cr}}$, where $\Sigma(\bm{\theta})$ is the projected surface mass density of the deflector and $\Sigma_{{\rm cr}}$ is the critical surface density for lensing given by
\begin{equation}
	\Sigma_{{\rm cr}} = \frac{c^2}{4\pi G} \frac{D_{{\rm s}}}{D_{{\rm d}}D_{{\rm ds}}},
\end{equation}
where $D_{{\rm s}}$, $D_{{\rm d}}$, and $D_{{\rm ds}}$ are the angular diameter distances between the observer and the source, between the observer and the deflector, and between the deflector and the source, respectively. The deflection potential is related to the convergence by the Poisson equation
\begin{equation}
	{\kappa} (\bm{\theta}) = \frac{1}{2} \bm{\nabla}_{\theta}^2 \psi( \bm{\theta} ).
	\end{equation}
The Einstein radius of the lens system is given by the solution of Equation \eqref{eq:lenseq} with $\beta = 0$ which is the case where the source lies directly behind the deflector. The Einstein radius can be expressed as
\begin{equation}
	\theta_{{\rm Ein}} = \sqrt{\frac{4GM(\theta_{{\rm Ein}}D_{{\rm d}})D_{{\rm ds}}}{c^2 D_{\rm s} D_{{\rm d}}}},
\end{equation}
where $M(r)$ is the enclosed mass of the deflector within a radius $r$. 

The time delay between two images is
\begin{equation}\label{eq:timedelay}
	\Delta t_{ij} = \frac{D_{\Delta t}}{c} \left[ \frac{1}{2} \left( \bm{\theta}_i - \bm{\beta}\right)^2 - \frac{1}{2} \left( \bm{\theta}_j - \bm{\beta}\right)^2  - \psi(\bm{\theta}_i) + \psi (\bm{\theta}_j) \right].
\end{equation}
Here $D_{\Delta t}$ is the time-delay distance given by
\begin{equation}
	D_{\Delta t} = (1+z_{\rm d}) \frac{D_{{\rm s}} D_{{\rm d}}}{D_{{\rm ds}}},
\end{equation}
where $z_{{\rm d}}$ is the redshift of the deflector.

The mass-sheet transformation \citep[MST;][]{FGS85} 
\begin{gather} \label{eq:mst}
	%\begin{gather}
	\kappa (\theta) \to \kappa'(\theta) = (1 - \lambda) + \lambda \kappa (\theta), \\
	\beta \to \beta' = \lambda \beta
	%\end{gather}
\end{gather}
leaves the image positions invariant. The additive term $(1-\lambda)$ can be internal to the deflector mass distribution affecting the time delay and the velocity dispersion as
\begin{equation}
	\begin{aligned}
	\Delta t' = \lambda \Delta t , \\
	\sigma_*' = \sqrt{\lambda} \sigma_*.
	\end{aligned}
\end{equation} 
Furthermore, this additive term can be due to the line-of-sight structures external to the deflector mass distribution, quantified as the external convergence $\kappa_{\rm ext} = 1-\lambda_{\rm ext}$, which only affects the time-delay. \citet{S+S13} point out that assuming a power-law profile for the deflector mass distribution breaks the MSD as the MST of a power law is not a power law. Therefore, it is necessary to consider more flexible models for the deflector mass distribution or families of mass models connected by the source-position transformation \citep[SPT;][]{SPT} to obtain unbiased measurements of the cosmological parameters.

\subsection{Deflector mass model}
\label{ssec:mass}

We need to model the mass distribution of the deflector in order to
compute spatially resolved kinematics of the deflector and lensing
data of the background source. We require this model to be realistic,
yet simple enough to be computationally efficient to create mock data
for numerous realizations of a lens system while performing the
Bayesian inference. Therefore, for simplicity we assume spherically
symmetric mass profiles for the deflector instead of elliptical mass
profiles. This assumption simplifies many computational tasks by
reducing a number of two-dimensional problems to only one-dimensional,
namely the radial, ones. Naturally, real lenses are typically not
spherical, so our spherical models are not intended literally, but to
be representative of non-spherical models, after marginalization over
all the non-spherical parameters. As we shall see in \autoref{sec:mock}
we will tune the uncertainties in our spherical models so as to effectively reproduce the uncertainty of non-spherical models.

Following standard practice \citep[e.g.][]{T+K02b,Suy++14}, we
describe the mass distribution of the deflector using two components:
dark matter and luminous matter, where the luminous matter resides
within a dark matter halo. We choose the Navarro-Frenk-White (NFW)
profile \citep{Navarro96} for the dark matter distribution and Jaffe
profile \citep{Jaffe83} for the luminous matter distribution. It is
empirically known that the total mass distribution in a galaxy, as a
combination of the dark matter and luminous matter distributions,
closely follows an isothermal profile, which is a power-law profile
with the power-law slope $\gamma\approx2$ \citep{T+K04, Koo++06,
  Koo++09, Auger10, D+T14}.

\subsubsection{NFW profile}

The NFW profile describes the mass distribution in the dark matter haloes as suggested by cosmological $N$-body simulations \citep{Navarro96, NFW97}. The spherical NFW profile has the form
\begin{equation}
	\rho(r) =  \frac{\rho_{{\rm s}}}{(r/r_{\rm s})(1+r/r_{{\rm s}})^2},
\end{equation}
where $\rho_{{\rm s}}$ and $r_{{\rm s}}$ are the scale density and radius, respectively. The convergence $\kappa$ implied by this mass profile is \citep{Bart96}
\begin{equation}
	\kappa (\theta) = \frac{2 \kappa_{{\rm s}}}{(x^2 - 1)} \left[ 1 - \mathcal{F}(x) \right],
	\end{equation}
where $x = \theta D_{{\rm d}}/r_{{\rm s}}$, $\kappa_{{\rm s}} = \rho_{{\rm s}} r_{{\rm s}}/\Sigma_{{\rm cr}}$ is the scale convergence, and the function $\mathcal{F}(x)$ is given by
\begin{equation} \label{eq:profilef}
	\mathcal{F}(x) =
	\left\{
		\begin{array}{ll}
			\sec^{-1} (x) / {\sqrt{x^2-1}}
			  & (x>1), \\
			1 & (x = 1), \\
			 {\rm sech}^{-1} {(x)} / {\sqrt{1-x^2}}  & (x<1). 
		\end{array}
	\right.
\end{equation}
The deflection angle for the NFW profile can be derived as \citep{MBM03}
\begin{equation}
	\alpha (\theta) = \frac{2}{\theta}\int^{\theta} \theta' \kappa(\theta') {\rm d}\theta' = \frac{4 \kappa_{{\rm s}} \theta_{{\rm s}} }{x} \left[ \ln(x/2) + \mathcal{F}(x) \right],
\end{equation}
where $\theta_{{\rm s}} = r_{{\rm s}} / D_{{\rm d}}$. The deflection potential for the NFW profile is then
\begin{equation}
	\psi (\theta) = \int \alpha(\theta) {\rm d}\theta = 2 \kappa_{{\rm s}} \theta_{{\rm s}}^2 \left[ \log^2 \left(\frac{x}{2}\right) + (x^2 -1) \mathcal{F}^2(x)\right].
\end{equation}

\subsubsection{Jaffe profile}
The Jaffe profile is given by
\begin{equation}
	\rho(r) =  \frac{\rho_{{\rm s}}}{(r/r_{{\rm s}})^2 (1+r/r_{{\rm s}})^2},
	\end{equation}
where $\rho_{{\rm s}}$ and $r_{{\rm s}}$ are the scale density and radius, respectively. This profile reproduces well the $R^{1/4}$ surface brightness profile in projection with $r_{{\rm s}}=R_{{\rm eff}}/0.763$, where $R_{{\rm eff}}$ is the effective radius. The convergence for the Jaffe profile is given by \citep{Jaffe83}
\begin{equation} \label{eq:kappa_jaffe}
	\kappa (\theta) = \kappa_{{\rm s}} \left[ \frac{\pi}{x} + 2 \frac{1 - (2-x^2)\mathcal{F}(x)}{1-x^2} \right], 
	\end{equation}
where $ x = \theta D_{{\rm d}}/r_{{\rm s}}$, $\kappa_{{\rm s}} = \rho_{{\rm s}} r_{{\rm s}}/\Sigma_{{\rm cr}}$ is the scale convergence, and $\mathcal{F}(x)$ is given in Equation \eqref{eq:profilef}. The deflection angle for the Jaffe profile can be derived as \citep{B+M04}
\begin{equation}
	\alpha (\theta) = 2 \kappa_{{\rm s}} \theta_{{\rm s}} \left[ \pi  - 2 x \mathcal{F}(x) \right],
	\end{equation}
where $\theta_{{\rm s}} = r_{{\rm s}}/D_{{\rm d}}$. The deflection potential that reproduces the convergence in Equation \eqref{eq:kappa_jaffe} is
\begin{equation}
	\psi (\theta) =2 \kappa_{{\rm s}} \theta_{{\rm s}}^2 \left[ \pi x + \log (x^2)  - 2(x^2-1) \mathcal{F}(x) \right].
\end{equation}

\subsubsection{Power-law mass profile}
The elliptical power-law model is often used to describe galaxy scale lenses \citep[e.g.][]{Suy++13}. In order to calibrate the uncertainty in our models we use the spherical power law mass density profile as a
baseline comparison. This mass density profile is given by
\begin{equation} \label{eq:powerlaw}
	\rho(r) = \rho_0 \left( \frac{r}{r_0} \right)^{-\gamma}.
\end{equation}
The deflection angle for the power law mass profile is given by
\begin{equation}
	\alpha (\theta) = \left( \frac{\theta_{{\rm Ein}}}{\theta} \right)^{\gamma-2} \theta_{\rm Ein},
\end{equation}
where $\theta_{\rm Ein}$ is the Einstein radius.

\section{Creating mock data}\label{sec:mock}

\begin{figure*}
	\includegraphics[width=2\columnwidth]{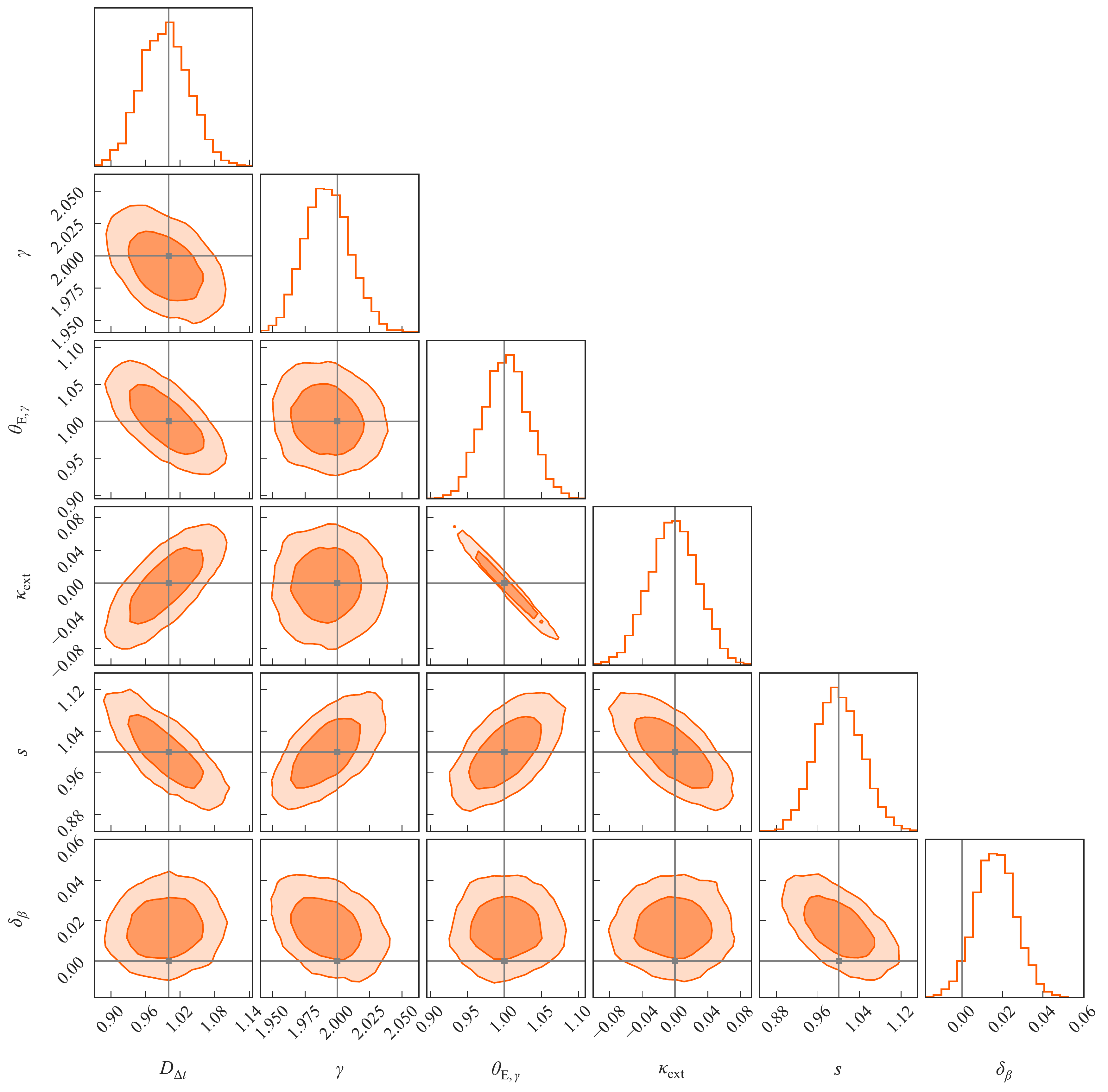}
	\caption{
	Posterior PDF of the model parameters for a power-law mass profile inferred from lensing data with 230 conjugate points. $D_{\Delta t}$, $\theta_{{\rm E}, \gamma}$, and $\delta_{\beta}$ are normalized with $D_{\Delta t}^{\rm fiducial}$, $\theta_{\rm E}$, and $\theta_{\rm E}$, respectively, where $\theta_{\rm E}$ is the true Einstein radius of the lens system. Grey lines show the true values of the parameters and orange contours show the 1$\sigma$ and 2$\sigma$ confidence regions. The uncertainty on the power-law slope is $\delta \gamma = 0.02$ and the time-delay distance $D_{\Delta t}$ is simultaneously estimated with 4.2 per cent uncertainty for an assumed Gaussian prior with 3 per cent uncertainty on $(1-\kappa_{\rm ext})$.
	\label{fig:conjugate_mc}
	}
\end{figure*}

\begin{figure}
	\includegraphics[width=\columnwidth]{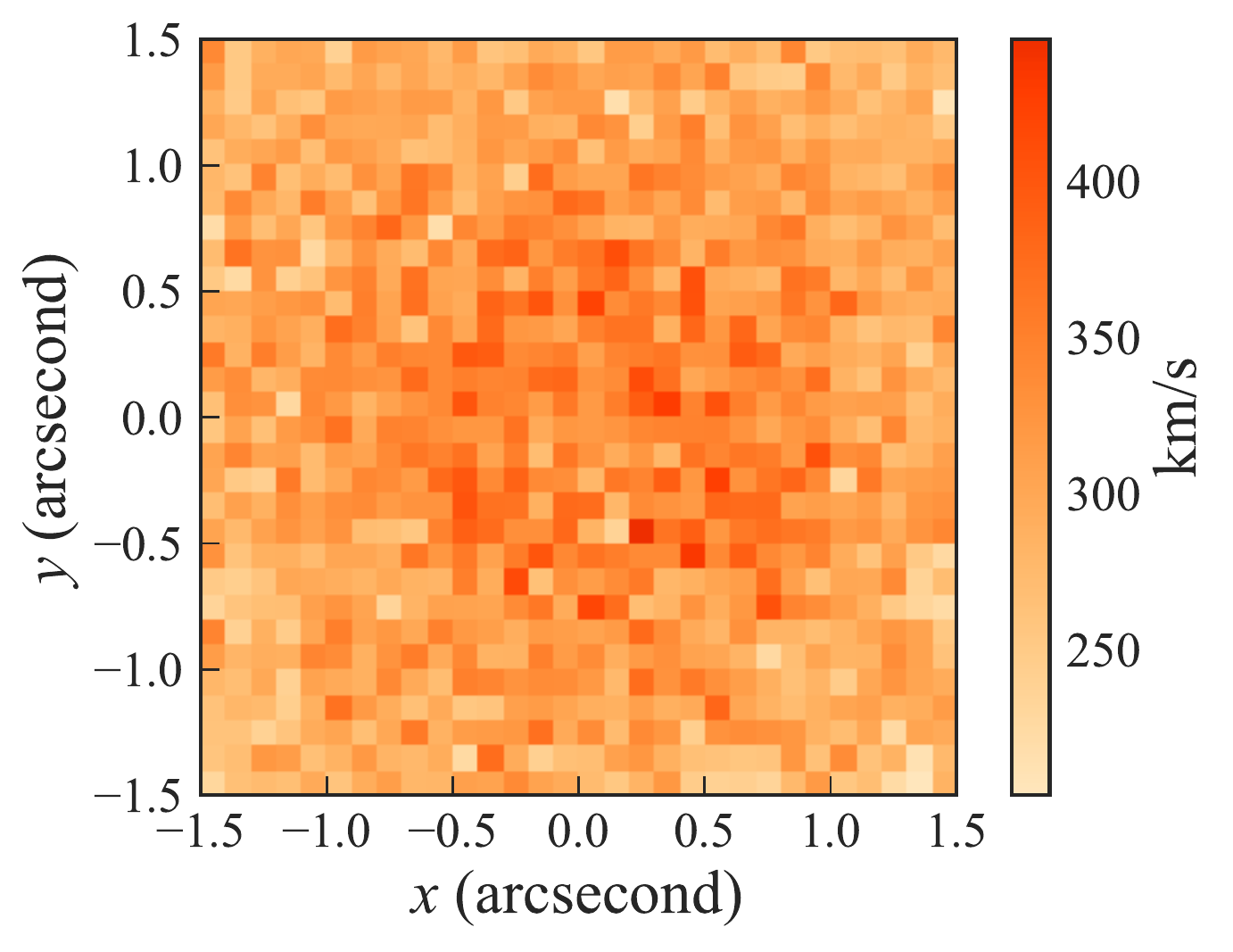}
	\caption{
		Line-of-sight velocity dispersion for a combination of NFW (dark component) and Jaffe (luminous component) profiles. 5 per cent random Gaussian noise was added to the velocity dispersion and it was smoothed with a Gaussian of FWHM=0.1 arcsec to take the effect of seeing into account.
	\label{fig:vel_dis_image}
	}
\end{figure}

In order to measure \Dd\ and \Ddt, three sets of data are necessary:
(1) imaging data of the lensed images of the quasar and its host
galaxy, (2) time delays from a monitoring campaign, and (3) kinematics
of the deflector. In this section we describe how we create mock data
of each kind for a given strong lens system. First, in
\autoref{ssec:conjugate_points} we describe how we use a set of
conjugate points to mimic the detailed modelling of the lensed quasar
host, which would be otherwise too computationally expensive to carry
out for large number of systems. Then, in \autoref{ssec:mock} we
describe how we create the full simulated data sets.

\subsection{Mimicking extended source reconstruction with conjugate points}\label{ssec:conjugate_points}
For the sake of speed, instead of carrying out a full extended source
reconstruction analysis, we describe each extended source as a set of
points, and analyse them with the so-called conjugate point techniques \citep{Gav++08}. In order to obtain realistic results, we need to determine how many
points to simulate and the associated astrometric uncertainty we want
to associate with each one. The amount of information depends on both
quantities, so we start by setting the latter and then adjust the
former to obtain a realistic precision. Computing time depends on
the number of points, so we adopt the smallest number that allows us
to achieve realistic precision on the model parameters while keeping
the computing time short enough for our purposes. In order to
calibrate our model we focus on the slope of the mass density profile
of a power-law mass model, which is the main parameter controlling the
velocity dispersion and time delay at a fixed \textit{Einstein} radius
\citep[e.g.][]{Wuc02,Suy12}. Thus, the minimum number of necessary source point is
chosen such that, for a power-law deflector mass profile given in
Equation \eqref{eq:powerlaw}, the power-law exponent can be inferred
from the set of conjugate points with an uncertainty
$\delta \gamma \sim0.02$.  We set this criterion to match with the precision on power-law slope $\gamma$ attainable by current
\citep{Suy++13,Wong17} and future technologies \citep{Meng++15} from a
full-blown computationally intensive lens modelling effort.

%e.g. the H0LiCOW project has achieved this level of
%precision for the lens HE0435-1223 \citep{Wong17}.
% could not find the appropriate paper for Meng et al 2016

We used a set of uniformly spaced points within a circle with 20 mas minimum separation between neighbouring points to mimic an extended source. We assumed a
power law mass profile given in Equation \eqref{eq:powerlaw} for the
deflector and created mock image data for the given source points. We
set the uncertainty in the image position as $\sigma_{\theta} = 60$
mas (corresponding to approximately half a pixel on the \textit{HST} Wide Field
Camera 3 infrared channel). The mock lens system in our analysis only produces two lensed images due to the assumed spherical symmetry. In doubly-imaged lens systems, there is a degeneracy between the power-law slope $\gamma$ and the Einstein radius $\theta_{{\rm Ein}}$ for asymmetric lens configurations whereas $\theta_{{\rm Ein}}$ is completely independent of $\gamma$ for a perfectly symmetric lens configuration \citep{Suy12}. Therefore, the number of conjugate points with fixed positional uncertainty needed to achieve a particular $\delta \gamma$ by breaking this degeneracy depends on the asymmetry of the lens configuration. We fix $\beta_{{\rm centre}} = \theta_{{\rm Ein}}/2$ for the rest of this study which is the case in the middle between the two extremes of perfect symmetry and maximal asymmetry.

We tuned this setup to give realistic errors on model parameters by
analysing mock data to obtain the posterior probability distribution function (PDF) of the model parameters: the power law slope of the mass profile $\gamma$, the
Einstein radius $\theta_{{\rm Ein}}$, and the source-point positions
$\bm{\beta}$. From Bayes' theorem, the posterior PDF follows
\begin{equation}
	P(\gamma, \theta_{{\rm Ein},\gamma}, \bm{\beta}, \kappa_{{\rm ext}} | \bm{\theta}) \propto P(\bm{\theta} | \gamma, \theta_{{\rm Ein},\gamma}, \bm{\beta}, \kappa_{{\rm ext}}) P(\gamma, \theta_{{\rm Ein},\gamma}, \bm{\beta}, \kappa_{{\rm ext}}),
\end{equation}
where $\bm{\theta}$ is the mock data for image positions, $\theta_{{\rm Ein},\gamma}$ is the Einstein radius for the power-law mass profile, and $\kappa_{\rm ext}$ is the external convergence. The first
term on the right-hand side is the likelihood of the data given the
model parameters, and the second one is the prior PDF of the model
parameters.

To sample from the posterior PDF through the Markov Chain Monte Carlo (MCMC) method, 
we use the \textsc{cosmoHammer} package \citep{Ake++12}, which embeds \textsc{emcee} \citep{Emcee13}, a \textsc{Python}
implementation of an affine-invariant ensemble sampler for MCMC
proposed by \mbox{\citet{G+W10}}. We first find the maxima of the likelihood function for the given image positions treating the source-point positions uncorrelated using the particle swarm optimization routine \citep{Kennedy95} included in \textsc{cosmoHammer}. We tuned the settings of the optimization process to find the maxima with $\sim$99 per cent accuracy. We then treat the source-point positions at the the maxima of the likelihood function as the reconstructed source. We sample from the posterior PDF of the source-point positions as
\begin{equation} \label{eq:source_sampling}
	\bm{\beta}_{{\rm sampled}} = s \bm{\beta}_{{\rm reconstructed}} + \delta_{\beta},
\end{equation} 
using two parameters: a rescaling factor $s$ for the source plane, and an offset $\delta_{\beta}$. Equation \eqref{eq:source_sampling} is essentially a SPT \citep{SPT}
\begin{equation}
	\bm{\beta} \to \bm{\beta}' = \left[ 1 + f(\bm{\beta}) \right] \bm{\beta},
\end{equation}
which is a generalization of the MST and leaves the strong lensing properties invariant. This allows us to incorporate the degeneracies induced by the SPT into our model.

We impose a Gaussian prior with 3 per cent uncertainty for $(1 - \kappa_{\rm ext})$ and uniform priors in appropriately large ranges for all the other parameters. The details of the chosen priors are given in \autoref{tab:conjugate_priors}.
\begin{table}
	\centering
	\caption{Priors for joint analysis with power-law mass profile}
	\label{tab:conjugate_priors}
	\begin{threeparttable}
	\begin{tabular}{cc} % four columns, alignment for each
		\hline
		Parameter &  Prior \\
		\hline
		$D_{\Delta t}$ & Uniform in [0, 2]$\times D_{\Delta t}^{{\rm fiducial}}$\tnote{*} \\
		$\gamma$ & Uniform in [1, 3] \\
		$\theta_{{\rm Ein},\gamma}$ & Uniform in [0.5, 2] arcsec\\
		$\kappa_{{\rm ext}}$ & Gaussian with 3 per cent uncertainty on $(1-\kappa_{\rm ext})$ \\
		$s$ & Uniform in [0, 2] \\
		$\delta_{\beta}$ & Uniform in [-0.5, 0.5] arcsec \\
		\hline
	\end{tabular}
	\begin{tablenotes}
	     \item[*] $D_{\Delta t}^{{\rm fiducial}}$ is the fiducial value of the time-delay distance.
	\end{tablenotes}
	\end{threeparttable}
\end{table}
After performing this analysis for various numbers of source points, we
find that the uncertainty of the power law exponent achieves our
target $\delta \gamma \sim 0.02$ for a source with 230 points
(\autoref{fig:conjugate_mc}). In comparison, a conservative choice of $\delta \gamma \sim 0.04$ can be achieved by adopting a source with 130 points. We also jointly sample the posterior PDF of the time-delay distance $D_{\Delta t}$ by adding a mock time-delay measurement to the data set. The posterior PDF of the joint analysis is
\begin{equation} \label{eq:joint_conjugate_likelihood}
	\begin{aligned}
	P(X | \bm{\theta}, \Delta t) &\propto P(\bm{\theta}, \Delta t | X) P(X) \\
	&\propto P(\bm{\theta} | X ) P(\Delta t | X ) P(X),
	\end{aligned}
\end{equation}
where $X$ are the model parameters $\{ D_{\Delta t}$, $\gamma$, $\theta_{{\rm Ein},\gamma}$, $\kappa_{{\rm ext}}$, $s$, $\delta_{\beta} \}$. The second line in Equation \eqref{eq:joint_conjugate_likelihood} is implied because the image positions and the time-delay data are independent measurements. The marginalized uncertainty of $D_{\Delta t}$ from the joint analysis is 4.2 per cent which is comparable to the state of the art measurements of the time-delay distance \citep{Suy++13, Wong17} after taking the difference in the uncertainty of $\kappa_{\rm ext}$ into account.
We thus conclude that the analysis of 230 correlated points with positional
uncertainty 60 mas with a spherical model approximates well the
extended source reconstruction with a non-spherical model as far as
the main parameters controlling \Dd\ and \Ddt\ are
concerned. Therefore, we adopt this setup when we analyse two
component mass models.

\subsection{Mock lensing data with spatially resolved velocity dispersion}
\label{ssec:mock}

We choose a composite mass model for the deflector galaxy assuming the NFW profile for the dark matter
component and the Jaffe profile for the luminous matter component.

%Like in every high dimensional space
%inference, it is key to choose the priors to be as informative as
%possible, as the volume can easily overtake the information content of
%the likelihood.

%As a realistic prior
We assumed that in projection one-third of the
total mass comes from the dark matter component within half of the
half-light radius \citep{Auger10}, to obtain the normalizations for
the NFW and Jaffe profiles.
%The normalization for the luminosity
%density profile $l(r)$ was obtained from assuming a mass-to-light
%ratio $M/L=10$ M${}_{\odot}$/L${}_{\odot}$.
%\textbf{ TT: why do you need this? You should not need the luminosity, no?}

First, we created mock lensing data for 230 conjugate points for the adopted deflector mass profile. Random Gaussian noise with standard deviation $\sigma_{\theta}=60$ mas was added to the conjugate point positions. 

The velocity dispersion profile for a mass distribution can be obtained by solving the spherical Jean's equation, which is given by
\begin{equation} \label{eq:jeans_eq}
	 \frac{1}{l(r)} \frac{{\rm d}(l \sigma_{r}^2)}{{\rm d}r} + 2 \beta_{{\rm ani}}(r) \frac{\sigma_r^2}{r} = - \frac{G M(\leq r)}{r^2}.
	\end{equation}
Here, $l(r)$ is the luminosity density of the galaxy, $\sigma_r$ is the radial velocity dispersion and $\beta_{{\rm ani}}(r)$ is the anisotropy profile given by
\begin{equation}
	\beta_{{\rm ani}} = 1 - \frac{\sigma_{{\rm t}}^2}{\sigma_r^2},
	\end{equation}
where $\sigma_{{\rm t}}$ is the tangential velocity dispersion for a spherically symmetric mass distribution. The surface-brightness-weighted, line-of-sight velocity dispersion can be obtained by solving this equation as \citep{Mamon05}
\begin{equation} \label{eq:los-vel-dis}
	I(R) \sigma^2_{{\rm los}}(R) = 2 G \int_R^{\infty} {\rm k} \left(\frac{r}{R}, \frac{r_{{\rm ani}}}{R} \right) l(r) M(r) \frac{{\rm d}r}{r}, 
\end{equation}
where $I(R)$ is the surface brightness. For Osipkov-Merritt anisotropy parameter $\beta_{{\rm ani}}(r) = 1/(1+r_{{\rm ani}}^2/r^2)$ \citep{Osipkov79, Merritt85a, Merritt85b}, the function ${\rm k}(u, u_{{\rm ani}})$ is given by
\begin{equation}
	\begin{aligned}
	{\rm k} (u, u_{{\rm ani}}) = & \frac{u_{{\rm ani}}^2+1/2}{(u_{{\rm ani}}^2 + 1)^{3/2}}\left( \frac{u^2+u_{\textrm {ani}}^2}{u} \right) \tan^{-1} \sqrt{\frac{u^2-1}{u_{\textrm {ani}}^2+1}} \\
	& \hspace{.1\textwidth} - \frac{1/2}{u_{{\rm  ani}}^2+1} \sqrt{1-1/u^2}.
	\end{aligned} 
	\end{equation}
Using Equation \eqref{eq:los-vel-dis}, we computed the line-of-sight velocity dispersion weighted by surface brightness in a given bin size (e.g. 0.1 arcsec). To take seeing into account, we convolved the surface-brightness-weighted line-of-sight velocity dispersion image with a Gaussian kernel of a given full width at half-maximum (FWHM) and then normalized it to obtain the line-of-sight velocity dispersion as
\begin{equation}
	\tilde{\sigma}^2_{{\rm los}} (x, y) = \frac{ I\sigma^2_{{\rm los}} \ast g (x,y)}{I \ast g(x,y)},
\end{equation}
where $g(x,y)$ is a two-dimensional Gaussian function, and the symbol `$\ast$' denotes the convolution (\autoref{fig:vel_dis_image}). Finally, we added random Gaussian noise with a given standard deviation to each pixel.
We also added a random Gaussian noise with 2 per cent standard deviation to
the time delay, typical of the best measurements \citep[e.g.][]{Bonvin17}.

\section{Precision on Cosmological Distance Measurements}\label{sect:distuncertainties}
\begin{figure*}
	\includegraphics[width=\columnwidth]{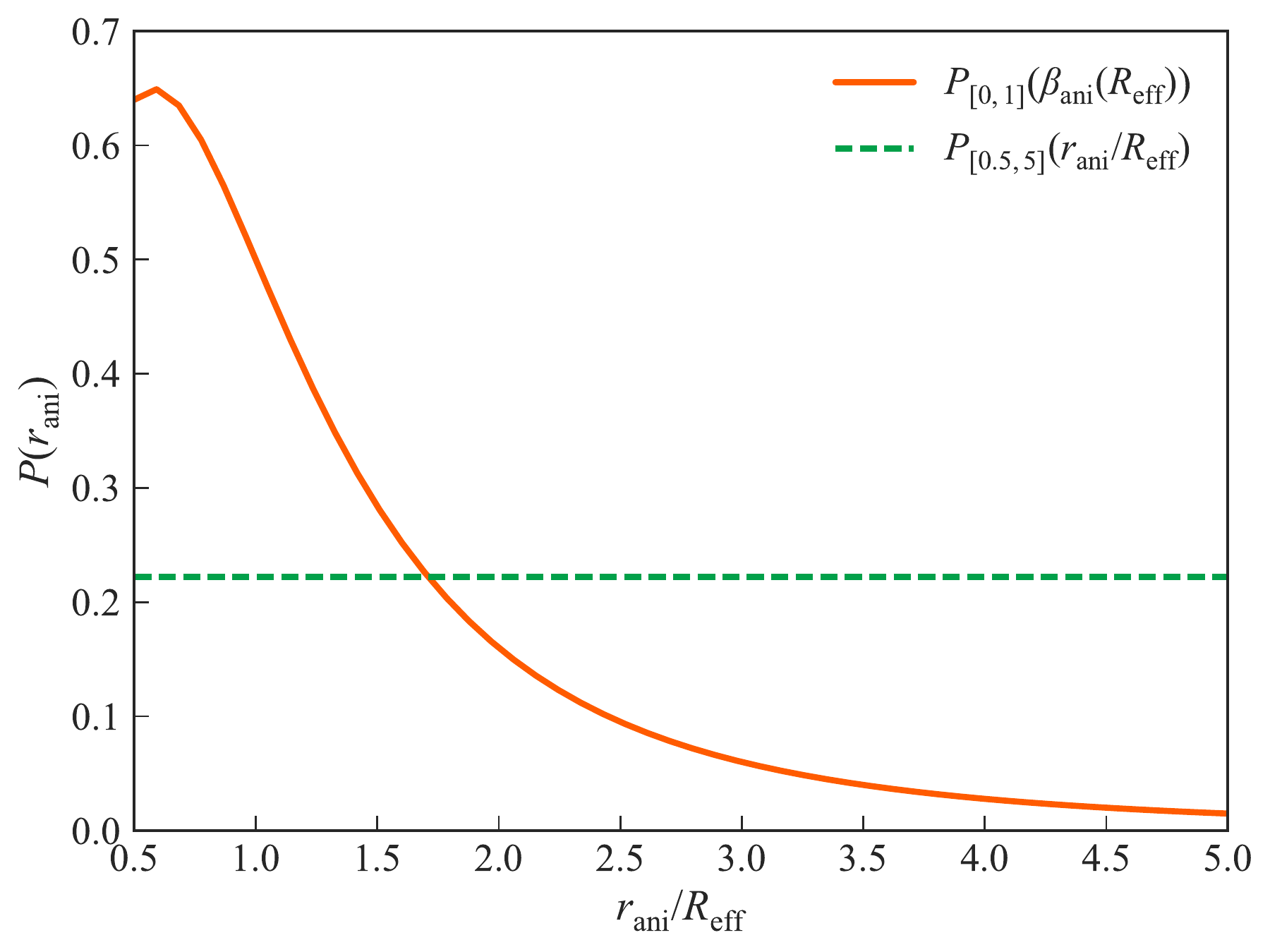}
	\includegraphics[width=\columnwidth]{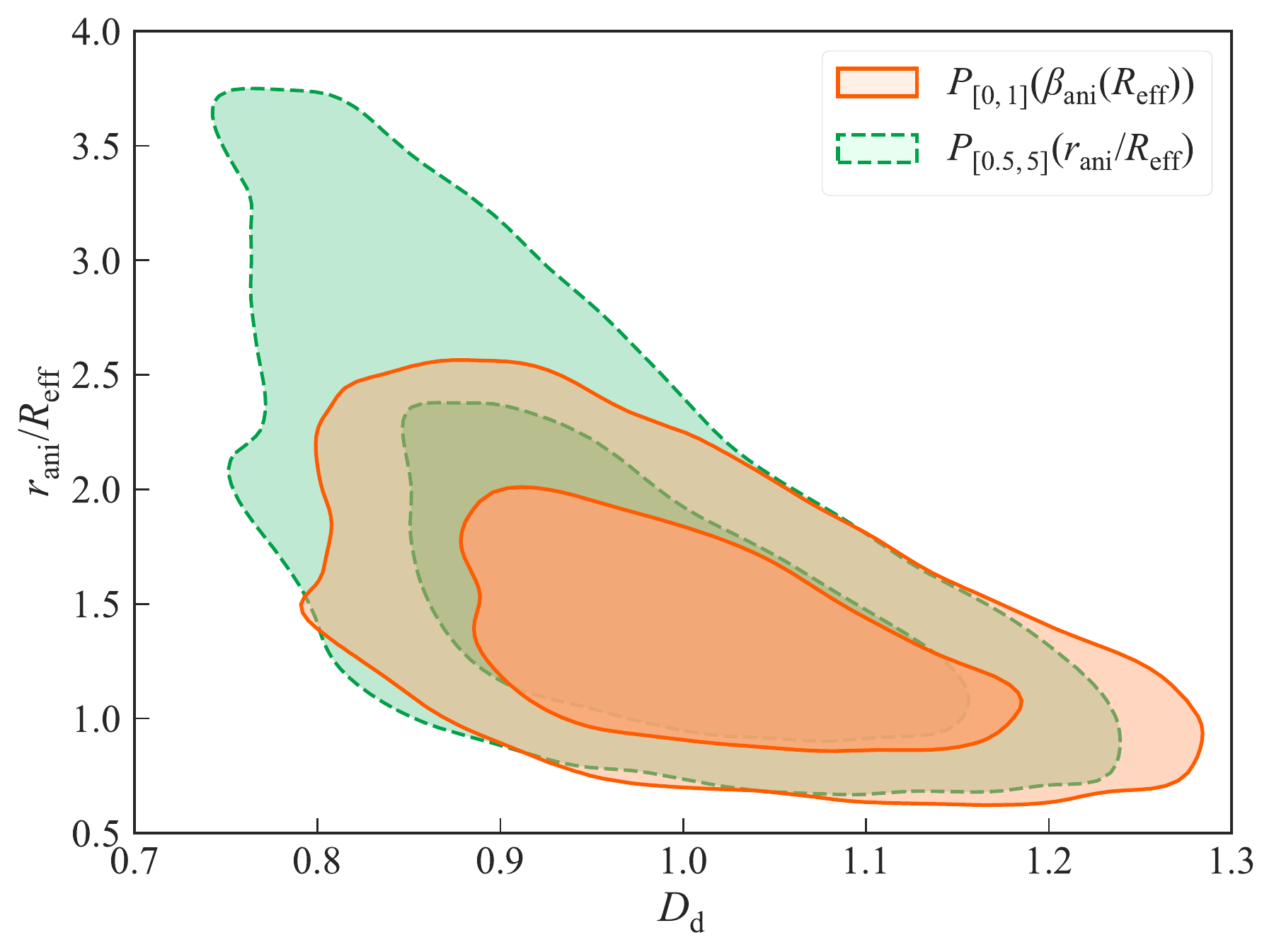}
	\caption{Priors for the anisotropy radius $r_{\rm ani}$ (left) and their effects on the mass-anisotropy degeneracy breaking (right). The two chosen priors are uniform prior for $r_{\rm ani}$ in [0.5, 5]$\times R_{{\rm eff}}$ [labelled $P_{[0.5, 5]}(r_{\rm ani}/R_{\rm eff})$, solid] and uniform prior for $\beta_{\rm ani} (R_{\rm eff})$ in [0, 1] [labelled $P_{[0,1]}(\beta_{\rm ani}(R_{\rm eff}))$, dashed]. In the right plot, the contours represent 1$\sigma$ and 2$\sigma$ confidence regions and $D_{\rm d}$ is normalized with $D_{\rm d}^{\rm fiducial}$. $P_{[0,1]}(\beta_{\rm ani}(R_{\rm eff}))$ puts more weight in the region $r_{\rm ani}/R_{\rm eff}<2$, where the assumed value of $r_{\rm ani}$ in our model lies, in comparison with $P_{[0.5, 5]}(r_{\rm ani}/R_{\rm eff})$ and it leads to a more unbiased and constrained estimate of the angular diameter distance $D_{\rm d}$.
	\label{fig:anisotropy_priors}
	}
	\end{figure*}
\begin{figure*}
	\includegraphics[width=2\columnwidth]{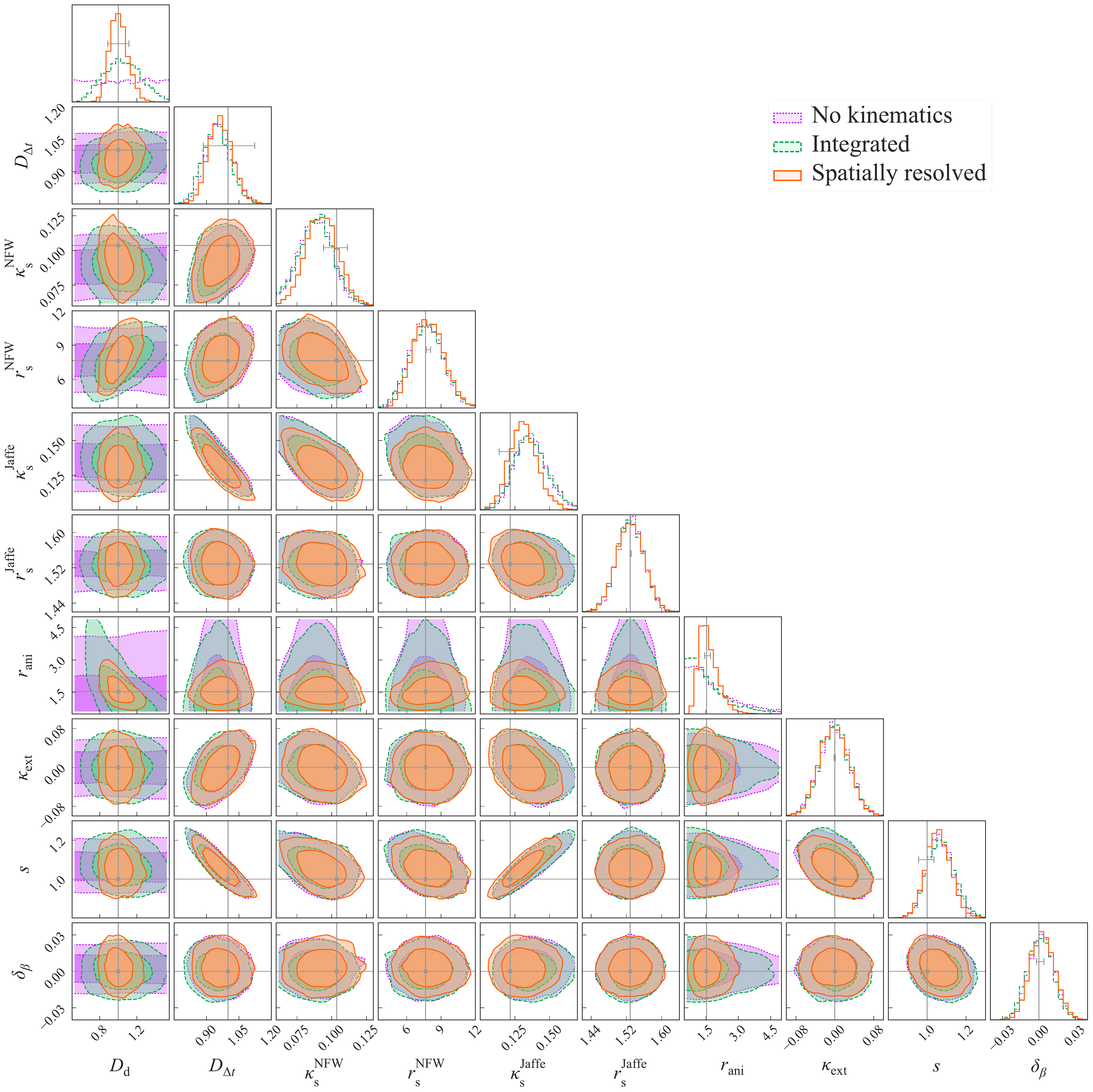}
	\caption{
	Posterior PDF of the model parameters given from joint analysis with lensing and time delay data with spatially resolved kinematics (solid), with integrated kinematics (dashed), and without any kinematics (dotted). The contours for each case represent 1$\sigma$ and 2$\sigma$ confidence regions. The model parameters $D_{{\rm d}}$, $D_{\Delta t}$, $r^{{\rm NFW}}_{\rm s}$, $r^{{\rm Jaffe}}_{\rm s}$, $r_{\rm ani}$, and $\delta_{\beta}$ are normalized with $D_{{\rm d}}^{{\rm fiducial}}$, $D_{\Delta t}^{{\rm fiducial}}$, $R_{{\rm Ein}}$, $R_{{\rm Ein}}$, $R_{{\rm Ein}}$, and $\theta_{{\rm Ein}}$, respectively, where $R_{{\rm Ein}}$ is the true Einstein radius with the dimension of length. Grey solid lines show the true values of the parameters. $D_{{\rm d}}$ can only be determined with kinematics. The anisotropy radius $r_{\rm ani}$ is also well determined with kinematics showing that the mass-anisotropy degeneracy is overcome.
	\label{fig:kinematics_corner}
	}
\end{figure*}
\begin{figure*}
	\includegraphics[width=2\columnwidth]{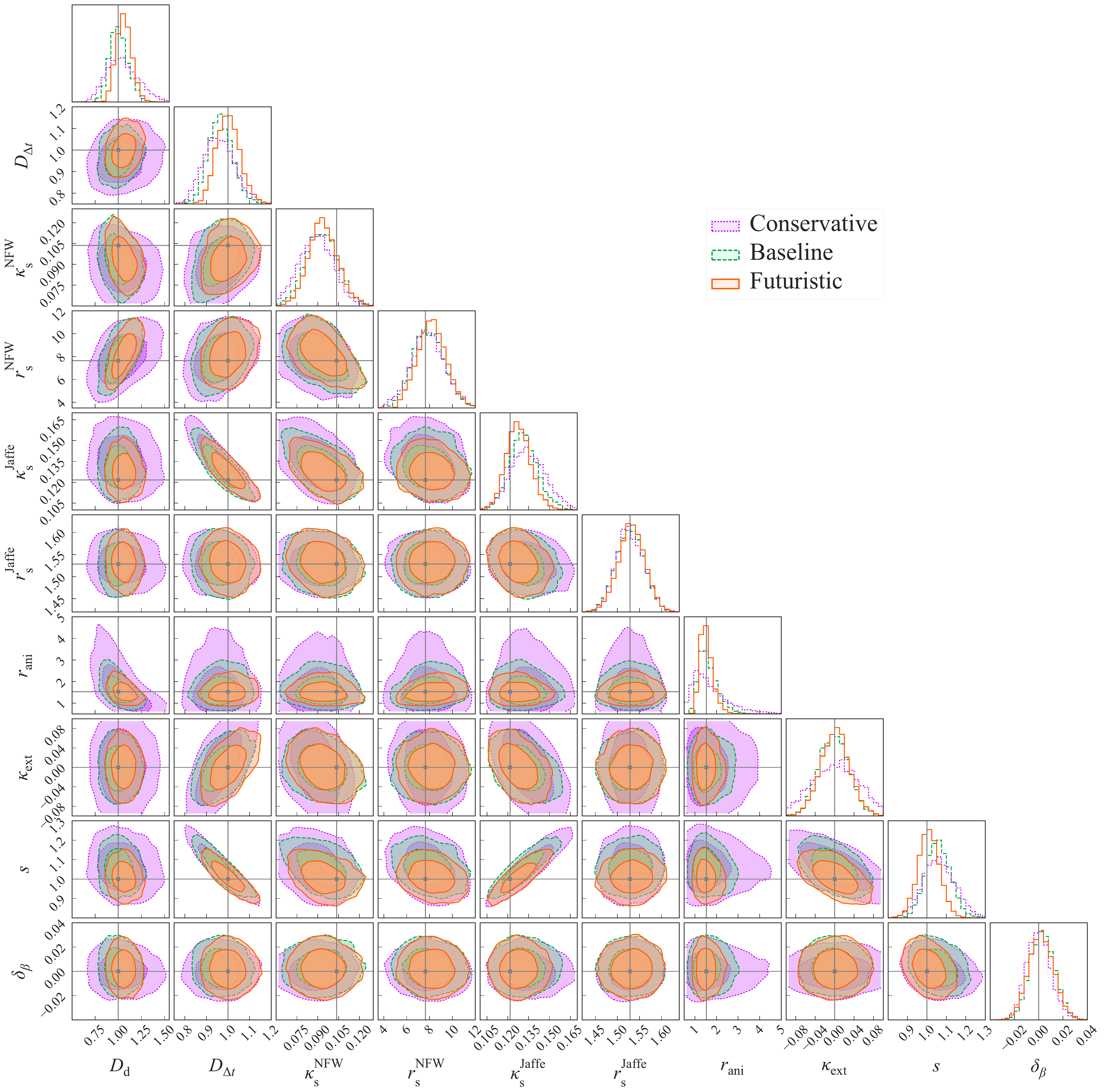}
	\caption{
	Posterior PDF of the model parameters given from joint analysis with lensing and time delay data with spatially resolved kinematics for baseline (dashed), futuristic (solid), and conservative (dotted) setups. The contours for each case represent 1$\sigma$ and 2$\sigma$ confidence regions. The model parameters $D_{{\rm d}}$, $D_{\Delta t}$, $r^{{\rm NFW}}_{\rm s}$, $r^{{\rm Jaffe}}_{\rm s}$, $r_{\rm ani}$, and $\delta_{\beta}$ are normalized with $D_{{\rm d}}^{\rm fiducial}$, $D_{\Delta t}^{\rm fiducial}$, $R_{{\rm Ein}}$, $R_{{\rm Ein}}$, $R_{{\rm Ein}}$, and $\theta_{{\rm Ein}}$, respectively, where $R_{{\rm Ein}}$ is the true Einstein radius with the dimension of length. Grey solid straight lines show the true values of the parameters. The constraints on the model parameters become tighter with higher quality of spatially resolved stellar kinematics.
	\label{fig:model_corner}
	}
\end{figure*}
In this section we use the mock data created as described in the
previous section to estimate the uncertainties of the angular diameter
and time-delay distances using the MCMC method.

We performed a joint analysis to obtain the posterior PDF of the model
parameters $X$ given the mock lensing data with velocity
dispersion and the time delay data. From Bayes' theorem, the posterior PDF
follows
\begin{equation}
	P(X | \bm{\theta}, \bm{\sigma}_{*}, \Delta t) \propto P(\bm{\theta}, \bm{\sigma}_{*}, \Delta t | X ) P(X),
\end{equation}
where $P(\bm{\theta}, \bm{\sigma}_{*}, \Delta t \left| X
\right.)$ is the likelihood of the data given the model parameters,
$P(X)$ is the prior PDF of the model parameters,	$\bm{\theta}$
is the image position data, $\bm{\sigma}_*$ is the velocity dispersion
data, $\Delta t$ is the time delay between images, and $X$
contains all the model parameters $\{ D_{{\rm d}},\ D_{\Delta t},\
\kappa_{\rm s}^{\rm NFW},\ r_{\rm s}^{{\rm NFW}},\ \kappa_{\rm s}^{\rm Jaffe},\
r_{\rm s}^{{\rm Jaffe}},\ r_{\rm ani},\ \kappa_{{\rm ext}},\ \bm{\beta}
\}$. As the image positions, the velocity dispersion, and the time
delay are independent data, the likelihood of the data given the model
parameters can be written as
\begin{equation}
 P(\bm{\theta}, \bm{\sigma}_{*}, \Delta t | X ) = P(\bm{\theta}|X) P (\bm{\sigma}_* | X) P(\Delta t | X).
\end{equation}
As it is often the case in high dimensional
spaces, it is important to choose the priors carefully
\citep[e.g.][]{Bre++14}. If the priors are not carefully chosen, the
marginalized one-dimensional posteriors on each parameter can be significantly
skewed \citep[e.g.][]{BAR16}, resulting in the median and mode of the
PDF to be a biased estimator of the true value. Naturally this bias
can be mitigated or eliminated by using the full PDF and not just
point estimators. However, it is important to use priors that are as
informative as possible. We impose Gaussian priors on $r_{{\rm s}}^{{\rm Jaffe}}$ and
$\kappa_{{\rm ext}}$, as $r_{{\rm s}}^{{\rm Jaffe}}$ can be measured
directly by fitting the surface brightness profile of the lens, whereas
$\kappa_{\rm ext}$ can be inferred indirectly by comparing the
statistics of galaxies along the line of sight to the lens with
simulated light cones \citep{Hil++09, Suy++13,Gre++13,Col++13,Rus++16}. We set a Gaussian prior for $r_{{\rm s}}^{{\rm NFW}}$ with 20 per cent uncertainty (\autoref{tab:joint_priors}). Note, this is a conservative choice comparing to the 14 per cent uncertainty adopted by \citet{Wong17} based on the results of \citet{Gav++07}.  We choose Jeffrey's prior $P(\xi) \propto 1/\xi$ for $\kappa_{\rm s}^{\rm NFW}$ and $\kappa_{\rm s}^{\rm Jaffe}$. We tested two prior choices for $r_{{\rm ani}}$: (a) uniform in [0.5, 5]$\times R_{{\rm eff}}$ \citep[hereafter referred to as $P_{[0.5, 5]}(r_{\rm ani}/R_{\rm eff})$, as used in][] {Suy++12, BAR16, Wong17}, and
(b) a uniform prior for $\beta_{\rm ani} (R_{\rm eff})$ in [0, 1] (hereafter referred as $P_{[0,1]}(\beta_{\rm ani}(R_{\rm eff}))$). $P_{[0,1]}(\beta_{\rm ani}(R_{\rm eff}))$ puts more weight in the region $r_{\rm ani}/R_{\rm eff}<2$, where the assumed value of $r_{\rm ani}$ in our model lies, in comparison with $P_{[0.5, 5]}(r_{\rm ani}/R_{\rm eff})$ and it results in a more unbiased and constrained estimate of the angular diameter distance $D_{\rm d}$ (\autoref{fig:anisotropy_priors}). Adopting a more restricting uniform prior for $r_{\rm ani}$ in $[0.5,2]\times R_{\rm eff}$ produces a similar constraint on $D_{\rm d}$ as the one by adopting $P_{[0,1]}(\beta_{\rm ani}(R_{\rm eff}))$. We set $P_{[0,1]}(\beta_{\rm ani}(R_{\rm eff}))$ as the prior for $r_{\rm ani}$ for the rest of this study.

\begin{table}
	\centering
	\caption{Priors for joint analysis with composite mass model}
	\label{tab:joint_priors}
	\begin{threeparttable}
	\begin{tabular}{cc} % four columns, alignment for each
		\hline
		Parameter &  Prior \\
		\hline
		$D_{{\rm d}}$ & Uniform in [0, 2]$\times D_{{\rm d}}^{{\rm fiducial}}$ \tnote{*} \\
		$D_{\Delta t}$ & Uniform in [0, 2]$\times D_{\Delta t}^{{\rm fiducial}}$\tnote{*} \\
		$\kappa_{\rm s}^{\rm NFW}$ & Jeffrey's prior \\
		$r_{{\rm s}}^{\rm NFW}$ & Gaussian with 20 per cent uncertainty \\
		$\kappa_{\rm s}^{\rm Jaffe}$ & Jeffrey's prior \\
		$r_{\rm s}^{\rm Jaffe}$ & Gaussian with 2 per cent uncertainty  \\
		$r_{{\rm ani}}$ & Uniform prior for $\beta_{\rm ani}$ in [0, 1] \\
		$\kappa_{{\rm ext}}$ & Gaussian prior on $(1-\kappa_{\rm ext})$ \\
		$s$ & Uniform in [0, 2] \\
		$\delta_{\beta}$ & Uniform in [-0.5, 0.5] arcsec \\
		\hline
	\end{tabular}
	\begin{tablenotes}
	     \item[*] $D_{{\rm d}}^{{\rm fiducial}}$ and $D_{\Delta t}^{ {\rm fiducial}}$ are the fiducial values of the angular diameter distance to the deflector and the time-delay distance.
	\end{tablenotes}
	\end{threeparttable}
\end{table}

We have examined the effect of having spatially resolved velocity
dispersion data on the uncertainties of the model parameters by
studying three cases: (1) without any kinematics, (2) with integrated
velocity dispersion data of the deflector within 1.2 arcsec radius, and
(3) with spatially resolved velocity dispersion data. We adopted three
observational settings, which reflect variation in qualities of
observation instruments and conditions. These settings are (1)
``baseline'': representative of the resolution and precision that can
be achieved with integral field spectrographs (IFSs) on current and
upcoming instruments, e.g. OSIRIS on Keck or NIRSPEC on the \textit{JWST},
with the precision on the velocity dispersion and the external
convergence that can be expected in the best cases, (2)
``conservative'': same as baseline but with conservative precision on
the velocity dispersion and the external convergence, and (3)
``futuristic'': for IFSs on upcoming extremely large telescopes, e.g.
IRIS on Thirty Meter Telescope (TMT) (\autoref{tab:models}). It is
beyond the scope of this paper to estimate the amount of exposure time
required to meet these goals for each one of the instrumental setups
and to analyse the sources of systematic uncertainties. This
exploration is left for future work.

As expected, \Dd\ can only be measured by adding the stellar kinematic information to the lensing and time-delay data
(\autoref{fig:kinematics_corner}). When integrated stellar kinematics is added, the anisotropy radius $r_{\rm ani}$ is not constrained due to the mass-anisotropy degeneracy. Given our
parametrization and assumptions, \Dd\ absorbs most of the improvement after adding the integrated stellar kinematics,
since the precision of \Ddt\ is limited by the assumed priors on time
delay and external convergence. If one were to consider more flexible
models, the gain would
be even more significant, highlighting the importance of kinematics.
In the real world of course, having additional information is not only
helpful for improving the precision but also for checking for
systematics and improving the accuracy. Using spatially resolved velocity dispersion
data can improve uncertainty on $D_{{\rm d}}$ from $\sim$20 to $\sim$10 per cent for
the baseline setup and from $\sim$27 to $\sim$17 per cent for the conservative setup
with respect to using integrated velocity dispersion data (\autoref{fig:model_corner}). Moreover, the anisotropy radius $r_{{\rm ani}}$ is well-determined only when spatially resolved kinematics is introduced
(\autoref{fig:kinematics_corner}), which demonstrates that spatially
resolved kinematics help break the mass-anisotropy degeneracy and allow
us to use the anisotropy radius $r_{{\rm ani}}$ as a free
parameter. For our adopted lensing data quality equivalent to $\delta \gamma \sim 0.02$, the lens model parameters are limited by modelling uncertainties, thus the addition of the spatially resolved kinematics improves the constraints only by $\sim$1 per cent. If we adopt a conservative lensing data quality equivalent to $\delta \gamma \sim 0.04$, the addition of the spatially resolved kinematics leads to more relative improvement in the constraints on the model parameters, e.g. uncertainty on $D_{\Delta t}$ improves by $\sim$3 per cent compared to the case with only integrated kinematics. In comparison to our adopted lensing data quality ($\delta \gamma \sim 0.02$), this conservative lensing data quality worsens the constraint $D_{\Delta t}$ by $\sim$2 per cent (from $\sim$6 to $\sim$8 per cent). The constraint on $D_{\rm d}$ does not significantly change (within 1 per cent), as $D_{\rm d}$ is limited by the quality of the stellar kinematics data. The uncertainties on $D_{{\rm d}}$ and $D_{\Delta t}$ for different
data sets and observational setups are summarized in Table
\ref{tab:uncertainties}.

To check for bias in point estimators of the model parameters, we
performed 25 joint analyses for different noise realizations using the same lensing parameters with the
``baseline'' setup. The 1$\sigma$ regions of the parameter estimates from these analyses
 are shown in with horizontal error bars in the one-dimensional histograms of \autoref{fig:kinematics_corner}. All the point estimators of the model parameters are within 1$\sigma$ of the true values. We note however, that it is
highly preferable to not adopt point estimators of individual
parameters, but rather take into account the full (asymmetric)
posterior PDF.

\begin{table*}
	\centering
	\caption{Parameters for different observational setups}
	\begin{threeparttable}
	\begin{tabular}{ccccccccc} % 9 columns, alignment for each
		\hline
		\multirow{2}{*}{Observational setup\tnote{*}} & \multirow{2}{*}{Annulus width} & \multirow{2}{*}{$N_{{\rm annuli}}$\tnote{$\dagger$}} &\multirow{2}{*}{PSF FWHM} & \multicolumn{5}{c}{Parameter uncertainties} \\
		& & & & ${a_{{\rm Jaffe}}}$ & ${1 - \kappa_{{\rm ext}}}$ & ${\Delta t}$ & $\sigma_{*}$ & $\bm{\theta}$  \\
		& (arcsecond) & & (arcsecond) & (per cent) & (per cent) & (per cent) & (per cent) & (mas)  \\
		\hline
		Baseline & 0.1 & 12 & 0.1 & 2 & 3 & 2 & 5 & 60  \\
		Conservative & 0.2 & 6 & 0.1 & 2 & 5 & 2 & 10 & 60  \\
		Futuristic & 0.05 & 24 & 0.03 & 2 & 3 & 2 & 5 & 60  \\
		\hline	\end{tabular}
	\label{tab:models}
	\begin{tablenotes}
	     \item[*] The ``baseline'' and ``conservative'' setups
               represent what we can expect to obtain with current and
               upcoming diffraction limited IFSs, e.g. OSIRIS on Keck and NIRSPEC on
               \textit{JWST}. The ``futuristic'' setup is for diffraction-limited IFSs on upcoming
               extremely large telescopes, e.g. TMT or E-ELT.
			 \item[$\dagger$] $N_{{\rm annuli}}$ refers to the number of annuli for the spatially resolved kinematics for each observational setup.
	\end{tablenotes}
  \end{threeparttable}
\end{table*}

\begin{table}
	\centering
	\caption{Uncertainties of $D_{{\rm d}}$ and $D_{\Delta t}$ for a single lens with different observational setups}
	\label{tab:uncertainties}
	\begin{tabular}{cccc} % four columns, alignment for each
		\hline
		Model & Kinematics data & $\sigma_{D_{\rm d}}$ & $\sigma_{D_{\Delta t}}$\\
		& & (per cent) & (per cent) \\
		\hline
		\multirow{3}{*}{Baseline} & No & - & 6.5  \\
		& Integrated &  19.8 & 6.5  \\
		& Resolved & 9.6 & 5.8 \\
		\hline
		\multirow{2}{*}{Conservative} & Integrated &  27.0 & 7.8  \\ 
		& Resolved & 16.7 & 7.5  \\
		\hline
		\multirow{1}{*}{Futuristic} & Resolved & 7.7  & 5.3  \\
		\hline
	\end{tabular}
\end{table}

\begin{table*}
	\centering
	\caption{Cosmological models and parameter priors}
	\label{tab:cosmo_models}
	\begin{tabular}{ccc} % four columns, alignment for each
		\hline
		Model name &  Description & Priors \\
		\hline
		$\Lambda$CDM & Flat $\Lambda$CDM cosmology & $h \in [0, 1.5]$, $\Omega_{{\rm m}} \in [0, 1]$ \\
		o$\Lambda$CDM & Non-flat $\Lambda$CDM cosmology & $h \in [0, 1.5]$, $\Omega_{\Lambda} \in [0, 1]$, $\Omega_{{\rm k}} \in [-0.5, 0.5]$, $\Omega_{{\rm m}} > 0$ \\
		$w$CDM & Flat $w$CDM cosmology & $h \in [0, 1.5]$, $\Omega_{\Lambda} \in [0, 1]$, $w \in [-2.5, 0.5]$ \\
		$N_{{\rm eff}}$CDM & Flat $N_{{\rm eff}}$CDM cosmology & $h \in [0, 1.5]$, $\Omega_{\Lambda} \in [0, 1]$, $N_{{\rm eff}} \in [1, 5]$ \\
		o$w$CDM & Non-flat $w$CDM cosmology & $h \in [0, 1.5]$, $\Omega_{\Lambda} \in [0, 1]$, $\Omega_{{\rm k}} \in [-0.5, 0.5]$, $\Omega_{{\rm m}} > 0$, $w \in [-2.5, 0.5]$ \\
		$w_a$CDM & Flat $w_a$CDM cosmology & $h \in [0, 1.5]$, $\Omega_{\Lambda} \in [0, 1]$, $w_0 \in [-2.5, 0.5]$, $w_a \in [-8, 4.5]$ \\
		\hline
	\end{tabular}
\end{table*}

\begin{table*}
	\centering
	\caption{Uncertainties of $D_{{\rm d}}$ and $D_{\Delta t}$ for different lens systems}
	\begin{threeparttable}
	\label{tab:lenses}
	\begin{tabular}{cccccc} % four columns, alignment for each
		\hline
		Lens system & $z_{{\rm d}}$ & $z_{{\rm s}}$ & Velocity dispersion data & $\sigma_{D_{\rm d}}$ & $\sigma_{D_{\Delta t}}$\\
		& & & & (per cent) & (per cent) \\
		\hline
		HE0047   &  0.41   &  1.66  & Resolved  &  9.6  &  5.9 \\
		J1206    &  0.75   &  1.79  & Resolved  &  8.8  &  5.4 \\
		HE0435   &  0.46   &  1.69  & Resolved  &  9.5  &  7.0 \\
		HE1104   &  0.73   &  2.32  & Resolved  &  9.1  &  5.5 \\
		RXJ1131  &  0.29   &  0.65  & Resolved  &  10.0  &  6.6 \\
		J0246    &  0.73   &  1.68  & Resolved  &  8.9  &  5.5 \\
		HS2209   &  0.38\tnote{*}   & 1.07 & Integrated   &  21.7  &  7.0 \\
		WFI2033  &  0.66   &  1.66  & Integrated  &  18.5  &  6.1 \\
		B1608    &  0.63   &  1.39  & Integrated  &  19.8  &  6.1 \\
		\hline
	\end{tabular}
	\begin{tablenotes}
	        \item [*] The deflector redshift for HS2209 has not been accurately measured yet, therefore we used a fiducial redshift of $z=0.38$. The results are not sensitive to the assumed redshift.
	      \end{tablenotes}
	\end{threeparttable}
\end{table*}

\section{Cosmological inference}
\label{sect:cosmoinfer}

Having estimated the precision attainable on the two distances for a
single lens, we now turn to the estimation of cosmological parameters
from samples of time-delay lenses.  First, in \autoref{ssec:alone}. we
investigate the precisions achievable from time delay lensing data
alone. Then, in \autoref{ssec:joint}, we combine the lensing
information with \textit{Planck} data to illustrate complementarity in
the determination of the cosmological parameters.

\subsection{Cosmology from strong lensing alone}
\label{ssec:alone}

We performed a Bayesian analysis to obtain the posterior PDF of the cosmological parameters ${C}$ given the inferred angular diameter and time-delay distances computed in \autoref{sect:distuncertainties}. The posterior PDF is given by Bayes' theorem as
\begin{equation}
	P({C} | {D}, Z) \propto P({D} | {C}, Z ) P({C}),
\end{equation}
where	${D}$ is the set of measurements of $D_{{\rm d}}$ and $D_{\Delta t}$ for the strong lenses, and $Z$ is the set of redshifts pairs $(z_{{\rm d}},\ z_{{\rm s}})$ for the lenses. To efficiently compute the likelihood term $P({D} | {C}, Z )$, we approximate the posterior PDF of $D_{{\rm d}}$ and $D_{\Delta t}$ of each lens by its best fit bivariate normal distribution function as
\begin{equation}\label{eq:gaussianfit}
	P(D_{{\rm d}}, D_{\Delta t}) = \frac{1}{2 \pi \sigma_{D_{{\rm d}}} \sigma_{D_{\Delta t}} \sqrt{1-\rho_{\rm cor}^2}} \exp \left[ - \frac{z(D_{\rm d}, D_{\Delta t})}{2(1-\rho_{\rm cor}^2)} \right],
	\end{equation}
where
\begin{equation}
	\begin{aligned}
	z(D_{\rm d}, D_{\Delta t}) = & \frac{\left(D_{{\rm d}} - \mu_{D_{{\rm d}}}\right)^2}{\sigma_{D_{{\rm d}}}^2} +  \frac{\left(D_{{\rm \Delta t}} - \mu_{D_{{\rm \Delta t}}}\right)^2}{ \sigma_{D_{{\rm \Delta t}}}^2} \\ 
	& - \frac{2 \rho_{\rm cor}(D_{\rm d} - \mu_{D_{\rm d}})(D_{\Delta t} - \mu_{D_{\Delta t}})}{\sigma_{D_{\rm d}}\sigma_{D_{\Delta t}}},
	\end{aligned}
	\end{equation}
and $\rho_{\rm cor} = {\rm cov} (D_{\rm d}, D_{\Delta t}) / \sigma_{D_{\rm d}}\sigma_{D_{\Delta t}}$ with ${\rm cov} (D_{\rm d}, D_{\Delta t} )$ being the covariance between the two distances. $\mu_{D_{{\rm d}}}$ and $\mu_{D_{{\rm \Delta t}}}$ are the means of $D_{{\rm d}}$ and $D_{\Delta t}$, respectively. Assuming the posterior PDF as a bivariate normal distribution function is accurate to the order of Fisher matrix approximation. As we are only interested in the precision of cosmological parameters, we choose $\mu_{D_{{\rm d}}}$ and $\mu_{D_{{\rm \Delta t}}}$ to be the fiducial values of the angular and time-delay distances.

The quoted uncertainties on the parameters are determined from the 16- and 84-percentiles of the posterior PDF. We have considered six different cosmological models for this analysis (\autoref{tab:cosmo_models}). The first one is the basic flat $\Lambda$CDM model. The next three models are one-parameter extensions of the basic $\Lambda$CDM model for $\Omega_K$, $w$, and $N_{{\rm eff}}$, labelled as o$\Lambda$CDM, $w$CDM, and $N_{{\rm eff}}$CDM models, respectively. The last two cosmological models are two-parameter extensions from the basic $\Lambda$CDM model, relaxing ($\Omega_{K}$, $w$) and ($w_0$, $w_a$), labelled as o$w$CDM and $w_a$CDM models, respectively. In the $w_a$CDM model, the dark-energy equation-of-state parameter $w$ is given by \citep{Chevallier01, Linder03}
\begin{equation}
	w(a) = w_0 + w_a (1-a),
\end{equation}
where, $a$ is the scale factor. We examined the parameter uncertainties primarily using the fiducial cosmology: $H_0 = 70$ km/s/Mpc, $\Omega_{{\rm m}} = 0.3$, $\Omega_{\Lambda} = 0.7$, $\Omega_K = 0$, $w = -1$.

%\begin{landscape}
\begin{table*}
	%\begin{sidewaystable*}
	\centering
			\caption{Uncertainties on cosmological parameters}
			\label{tab:cosmo_params}
			\begin{adjustbox}{width=\textwidth}
			\begin{threeparttable}
			\begin{tabular}{ccccccccccc} % four columns, alignment for each
				\hline
				Data sets & $H_0$ & $\sigma(H_0)$ & $\Omega_{{\rm m}}$ & $\sigma$($\Omega_{{\rm m}}$) & $\Omega_{{\rm k}}$ & $\sigma(\Omega_{{\rm k}})$ &  $w$ & $\sigma(w)$ & $N_{{\rm eff}}$ & $\sigma(N_{{\rm eff}})$ \\
				& (km/s/Mpc) & (per cent) & & & & & & & & \\
				\hline
				\multicolumn{11}{c}{$\Lambda$CDM} \\
				\hline
				L9\tnote{*}                 & $69.7 \pm 1.4$  &  2.0  &  $0.33^{+0.09}_{-0.12}$  & 0.11  & -  &  -  &  -  &  -        &  -  &  -\\
				L40\tnote{*}                & $69.91 \pm 0.64$ & 0.92   &  $0.307^{+0.042}_{-0.047}$ &  0.044  & -  &  -  &  -  &  -        &  -  &  -\\
				L9+\textit{Planck}          & $68.35^{+0.82}_{-0.73}$  & 1.1   & $0.300\pm0.010$  &  0.010 &  - &  -  & -   & -   &  -  &  -         \\ 
				L40+\textit{Planck}         & $69.45^{+0.59}_{-0.43}$   & 0.74  & $0.2866^{+0.0058}_{-0.0061}$  & 0.0059   &  - &  -  & -   & -        &  -  &  -    \\
				\hline
				\multicolumn{11}{c}{o$\Lambda$CDM} \\ 
				\hline
				L9                 & $69.6 \pm 2.3$   & 3.3   & $0.35^{+0.19}_{-0.22}$  & 0.20  & $-0.01 \pm 0.27$  &  0.27  & -  & -  &  -  &  -  \\
				L40                & $70.0 \pm 1.1$   & 1.6   & $0.308^{+0.087}_{-0.091}$  & 0.089  & $0.00\pm0.12$  & 0.12   &  -  & -   &  -  &  -  \\ 
				%L9                & $55.47^{+1.66}_{-1.61}$   & 2.9   & $0.45^{+0.19}_{-0.18}$  & 0.19  & $-0.02^{+0.32}_{-0.30}$  & 0.31   &  -  & -   &  -  &  -  \\
				%L40                 & $55.75^{+1.19}_{-1.20}$   & 2.1   & $0.43\pm0.12$  & 0.12  & $-0.03^{+0.27}_{-0.25}$  &  0.26  & -  & -  &  -  &  -  \\
				L9+\textit{Planck}\tnote{$\dagger$}  & $56.3^{+1.1}_{-0.9}$  & 1.8   & $0.443^{+0.018}_{-0.016}$ & 0.017  & $-0.0341 \pm 0.0048$  & 0.0048   & -   & -  &  -  &  -    \\
				L40+\textit{Planck}\tnote{$\dagger$} &  $56.47^{+0.44}_{-0.47}$  & 0.81  & $0.441^{+0.011}_{-0.009}$  & 0.010  & $-0.0337^{+0.0033}_{-0.0030}$  & 0.0031    &  -  &  -  &  -  &  -   \\ % omega_m = 0.4418 + 0.0124 - 0.0082
				\hline
				\multicolumn{11}{c}{$w$CDM} \\ 
				\hline 
				L9                 & $70.2^{+3.5}_{-4.3}$   & 6.2  & $0.34^{0.12}_{0.13}$  &  0.13 & - & -  & $-1.11^{+0.66}_{-0.48}$   &  0.57 &  -  &  -  \\ 
				L40                & $70.2^{+1.8}_{-2.2}$   & 2.9   & $0.307 \pm 0.050$  & 0.050  &  - &  -  & $-1.03^{+0.24}_{-0.21}$  & 0.22 &  -  &  -    \\ 
				L9+\textit{Planck}          & $71.9^{+2.1}_{-1.8}$   & 2.8   & $0.276^{+0.014}_{-0.015}$  & 0.015  & -   & -   & $-1.157\pm0.081$   & 0.081  &  -  &  -    \\ 
				L40+\textit{Planck}         & $71.50^{+0.96}_{-0.79}$   & 1.2   & $0.2779^{+0.0060}_{-0.0058}$  & 0.0059 & - & - & $-1.127^{+0.054}_{-0.067}$  &  0.060  &  -  &  -     \\ 
				
				\hline
				\multicolumn{11}{c}{$N_{{\rm eff}}$CDM} \\ 
				\hline 
				L9 & $69.7^{+1.4}_{-1.3}$ & 2.0 & $0.33^{+0.09}_{-0.13}$ & 0.11 & - & - & - & - &$ 3.0^{+1.3}_{-1.4}$ & 1.4 \\
				L40 & $69.94 \pm 0.65$ & 0.93 & $0.305^{+0.042}_{-0.047}$ & 0.045 & - & - & - & - & $ 3.0 \pm 1.4$ & 1.4 \\
				L9+\textit{Planck} & $69.7^{+1.1}_{-1.0}$ & 1.6 & $0.299^{+0.011}_{-0.010}$ & 0.010 & - & - & - & - &$ 3.31 \pm 0.16$ & 0.16 \\
				L40+\textit{Planck} & $69.94^{+0.55}_{-0.54}$ & 0.77 & $0.2971^{+0.0084}_{-0.0076}$ & 0.0080 & - & - & - & - & $3.33^{+0.13}_{-0.12}$ & 0.13 \\
				
				\hline
				\multicolumn{11}{c}{o$w$CDM} \\ 
				\hline 
				L9 & $71.1^{+4.0}_{-5.2}$ & 6.5 & $0.41^{+0.24}_{-0.20}$ & 0.22 & $-0.10^{+0.26}_{-0.29}$ & 0.28 & $-1.25^{+0.72}_{-0.55}$ & 0.63 & - & -  \\
				L40 & $70.7^{+2.0}_{-3.1}$ & 3.6 & $0.36^{+0.19}_{-0.14}$ & 0.17 & $-0.06^{+0.19}_{-0.22}$ & 0.21 & $-1.14^{+0.46}_{-0.38}$ & 0.42 & - & -  \\
				\hline
				\multicolumn{11}{c}{$w_a$CDM} \\
				\hline
				Data sets & $H_0$ & $\sigma(H_0)$ & $\Omega_{{\rm m}}$ & $\sigma$($\Omega_{{\rm m}}$) & $\Omega_{\Lambda}$ & $\sigma(\Omega_{\Lambda})$ &  $w_0$ & $\sigma(w_0)$ & $w_a$ & $\sigma(w_a)$ \\
				& (km/s/Mpc) & (per cent) & & & & & & & & \\
				\hline 
				L9 & $70.4^{+5.0}_{-5.8}$ & 7.7 & $0.40^{+0.12}_{-0.13}$ & 0.13 & $0.60_{-0.13}^{+0.12}$ & 0.13 & $-0.98^{+0.86}_{-0.77}$ & 0.82 & $ -2.2^{+3.7}_{-3.3}$ & 3.5 \\ 
				L40 & $68.7^{+3.7}_{-3.6}$ & 5.3 & $0.359^{+0.092}_{-0.078}$ & 0.085 & $0.641_{-0.078}^{+0.092}$ & 0.085 & $-0.77^{+0.46}_{-0.64}$ & 0.55 & $ -1.6^{+3.5}_{-2.4}$ & 3.0 \\
				L9+\textit{Planck}+BAO & $65.5^{+2.6}_{-2.2}$ & 3.7 & $0.335 \pm 0.025$ & 0.025 & $0.665\pm0.025$ & 0.025 & $-0.59^{+0.28}_{-0.29}$ & 0.29 & $ -1.46^{+0.86}_{-0.85}$ & 0.86 \\ 
				L40+\textit{Planck}+BAO & $67.0^{+2.2}_{-2.0}$ & 3.2 & $0.321^{+0.022}_{-0.020}$ & 0.021 & $0.679^{+0.020}_{-0.022}$ & 0.021 & $-0.67^{+0.23}_{-0.26}$ & 0.25 & $ -1.39^{+0.75}_{-0.77}$ & 0.76 \\
				
				%L9+\textit{Planck}+BAO+JLA & $68.61^{+1.03}_{-1.01}$ & 1.5 & $0.31\pm0.01$ & 0.01 &$ 0.69\pm0.01$ & 0.01 & $-0.88^{+0.12}_{-0.10}$ & 0.11 & $ -0.73^{+0.40}_{-0.46}$ & 0.43 \\
				%L40+\textit{Planck}+BAO+JLA & $69.50^{+0.97}_{-0.95}$ & 1.4 & $0.30\pm0.01$ & 0.01 & $0.70\pm0.01$ & 0.01 & $-0.77^{+0.08}_{-0.13}$ & 0.11 & $ -1.29^{+0.43}_{-0.47}$ & 0.45 \\
				\hline\\
				
			\end{tabular}
			\begin{tablenotes}
				\item[*] L9 refers to the set of nine lenses and L40 refers to the set of 40 lenses.
				\item[$\dagger$] For o$\Lambda$CDM model, to combine \textit{Planck} with the lensing information, the fiducial cosmology was chosen to be the \textit{Planck} o$\Lambda$CDM cosmology: $H_0 = 56.5$ km/s/Mpc, $\Omega_{{\rm m}} = 0.441$, $\Omega_{\Lambda} = 0.592$, and $\Omega_{K} = -0.033$.
			\end{tablenotes}
		\end{threeparttable}
	\end{adjustbox}
	%\end{sidewaystable*}
\end{table*}
%\end{landscape}

\begin{figure}
	\includegraphics[width=1\columnwidth]{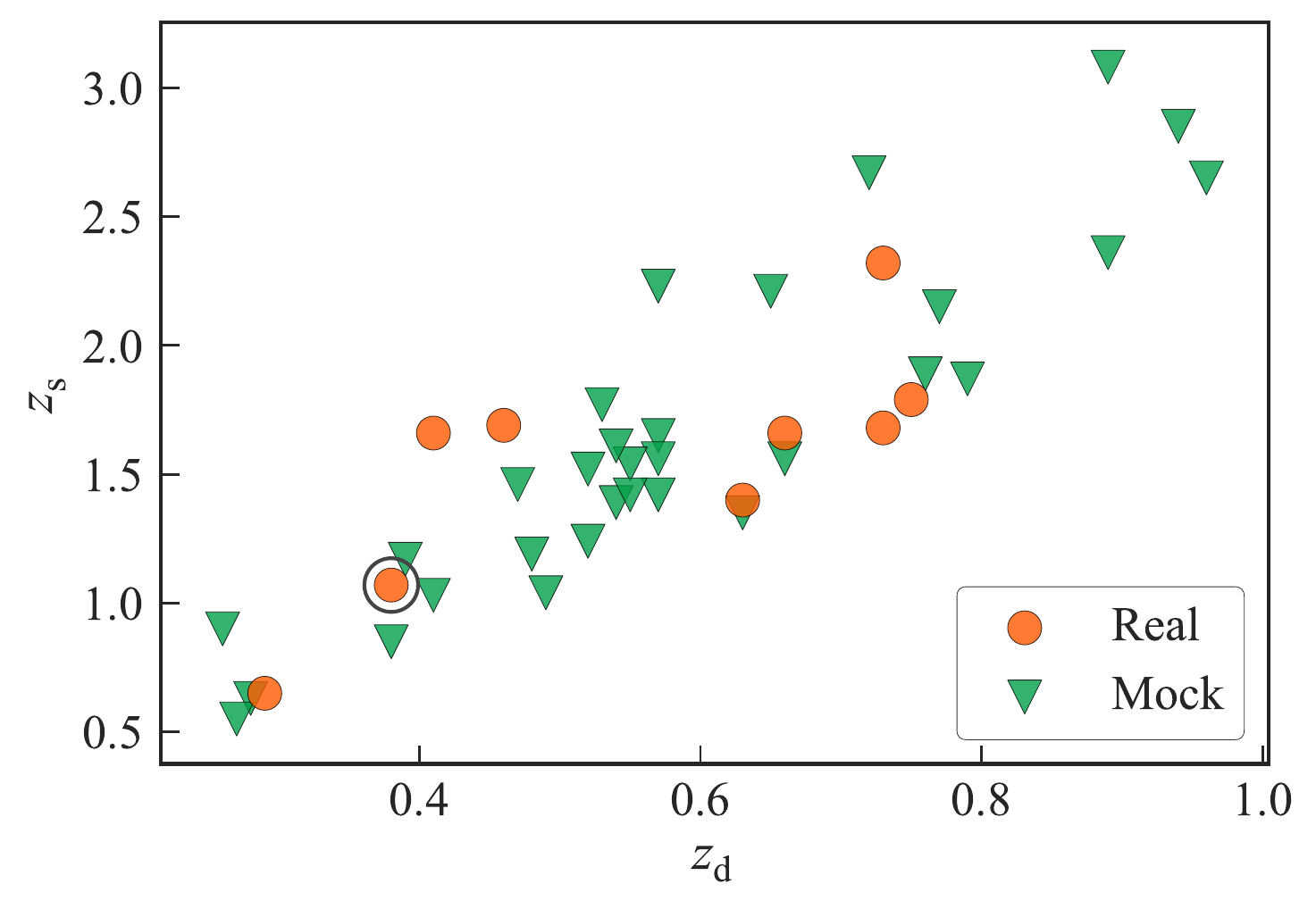}
	\caption{
	Distribution of deflector and source redshifts of the
        lenses. The circles show the redshifts for the nine actual
        lenses with measured time delays and deep \textit{HST} imaging. The triangles show the
        redshifts for the 31 lenses in the mock catalogue. We assume a fiducial redshift $z_{{\rm d}}=0.38$ for the strong lens HS2209 as it has not been accurately measured yet and it is marked with a dark circle on the plot.
	\label{fig:lens_zs}
	}
\end{figure}

First, we explored the uncertainties on the cosmological parameters
achievable by using nine real lenses for which accurate time delay
measurements and deep \textit{HST} imaging data are readily available. The
details of these nine lenses are given in \autoref{tab:lenses}.  Out
of these nine lenses, we consider six lenses to have spatially
resolved kinematics and the remaining three to have integrated
kinematics from the ``baseline'' observational setup, since three of
the lenses are currently outside of the reach of OSIRIS on
Keck. Spatially resolved kinematics for all nine systems could be
obtained with \textit{JWST}, so our estimate should be considered as
conservative in this respect. Then, to explore the strength of using
strong lenses to measure the cosmological parameters, we repeated the
analysis for a simulated sample of 40 strong lenses expected to be
available in the next few years through dedicated follow-up of newly
discovered systems. Thus, we created a mock catalogue of 31 lenses with
a redshift distribution that resembles the one for the nine lenses
given in \autoref{tab:lenses} in the following manner. First, we fit a Gaussian distribution to the redshift distribution of deflectors of the nine lenses and sampled from this fitted Gaussian distribution. Next, we also fit a Gaussian distribution to the distribution of the ratios of the deflector and source redshifts from the nine lenses and sampled from this distribution to determine the source redshift for each of the 31 mock lensing systems. The redshift distribution of the real and mock lenses is shown in \autoref{fig:lens_zs}.

\begin{figure} 
	\includegraphics[width=1\columnwidth]{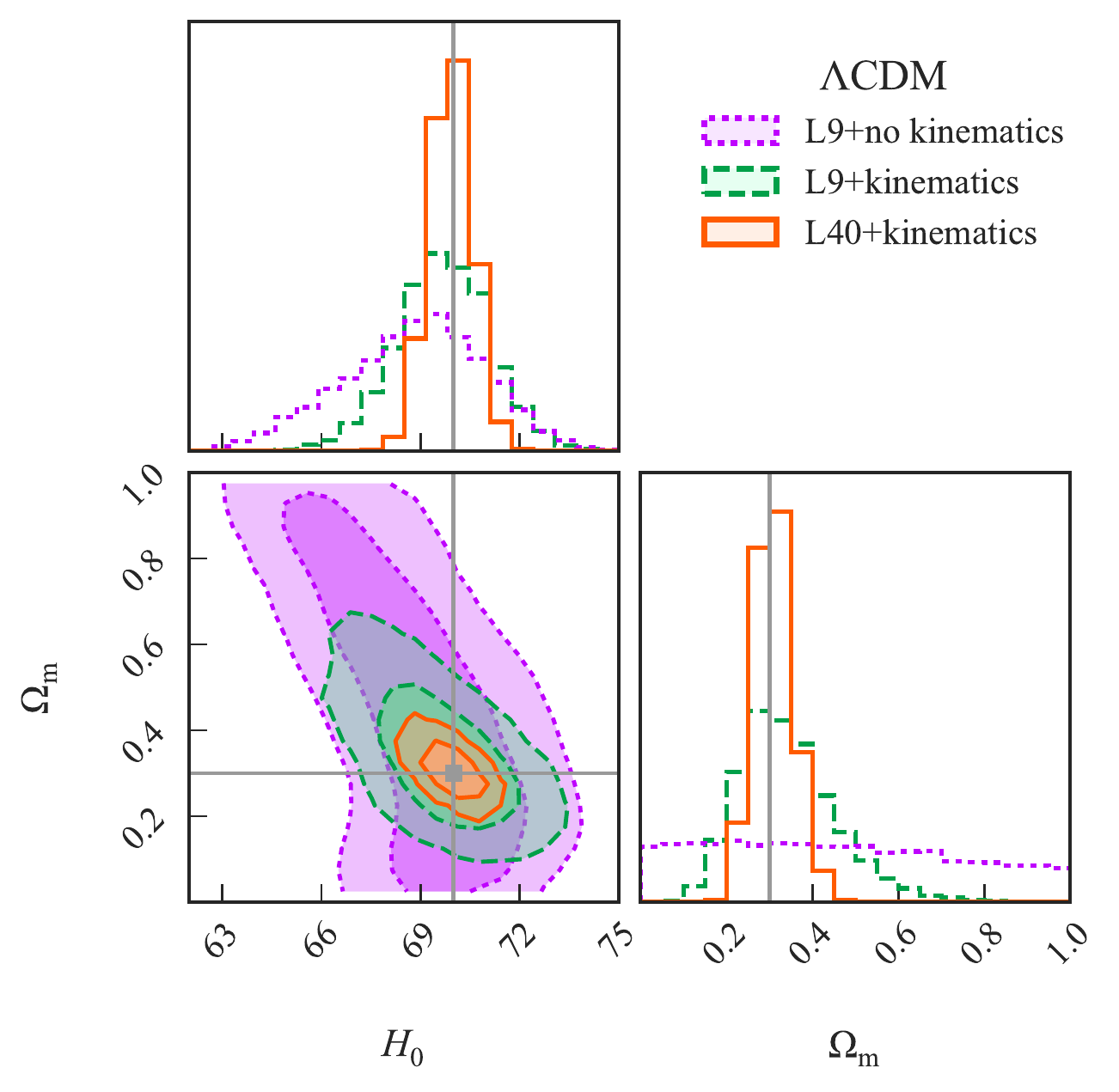} \\
	\caption{
	Posterior PDF of cosmological parameters for the flat $\Lambda$CDM
        model obtained from distance measurements for nine lenses (L9)
        without kinematics (dotted), for nine lenses with kinematics
        (dashed) and for 40 lenses (L40) with kinematics (solid). The contours represent 1$\sigma$ and 2$\sigma$ confidence regions. Solid straight lines
        show the fiducial values. Using kinematics breaks the
        degeneracy between parameters and improves the precision on
        $H_0$  from 3.2 to 2.0 per cent for nine lenses.
	\label{fig:LCDM_9lens}
	}
\end{figure}

\begin{figure*}
	\includegraphics[width=1\columnwidth]{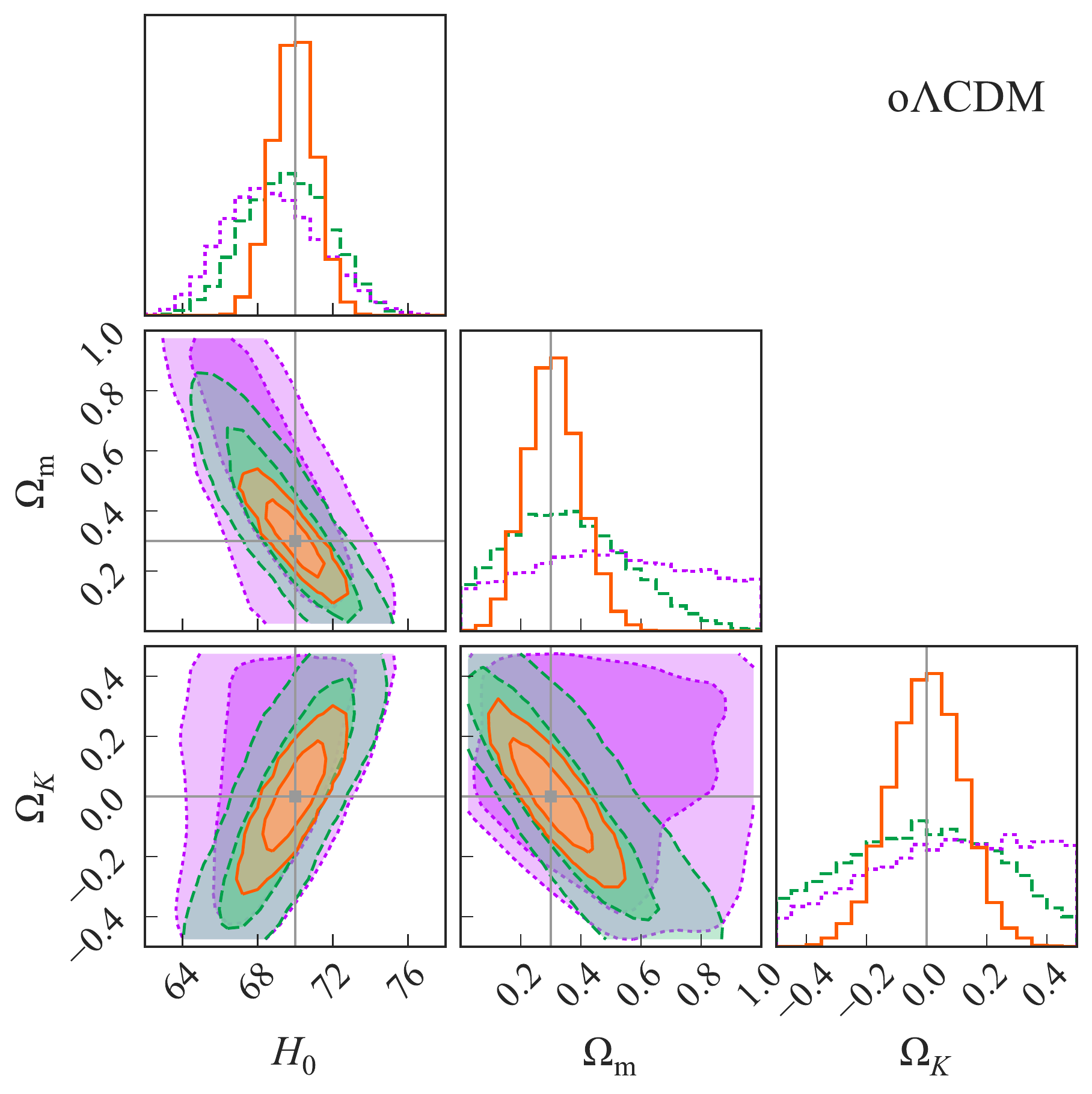}
	\includegraphics[width=1\columnwidth]{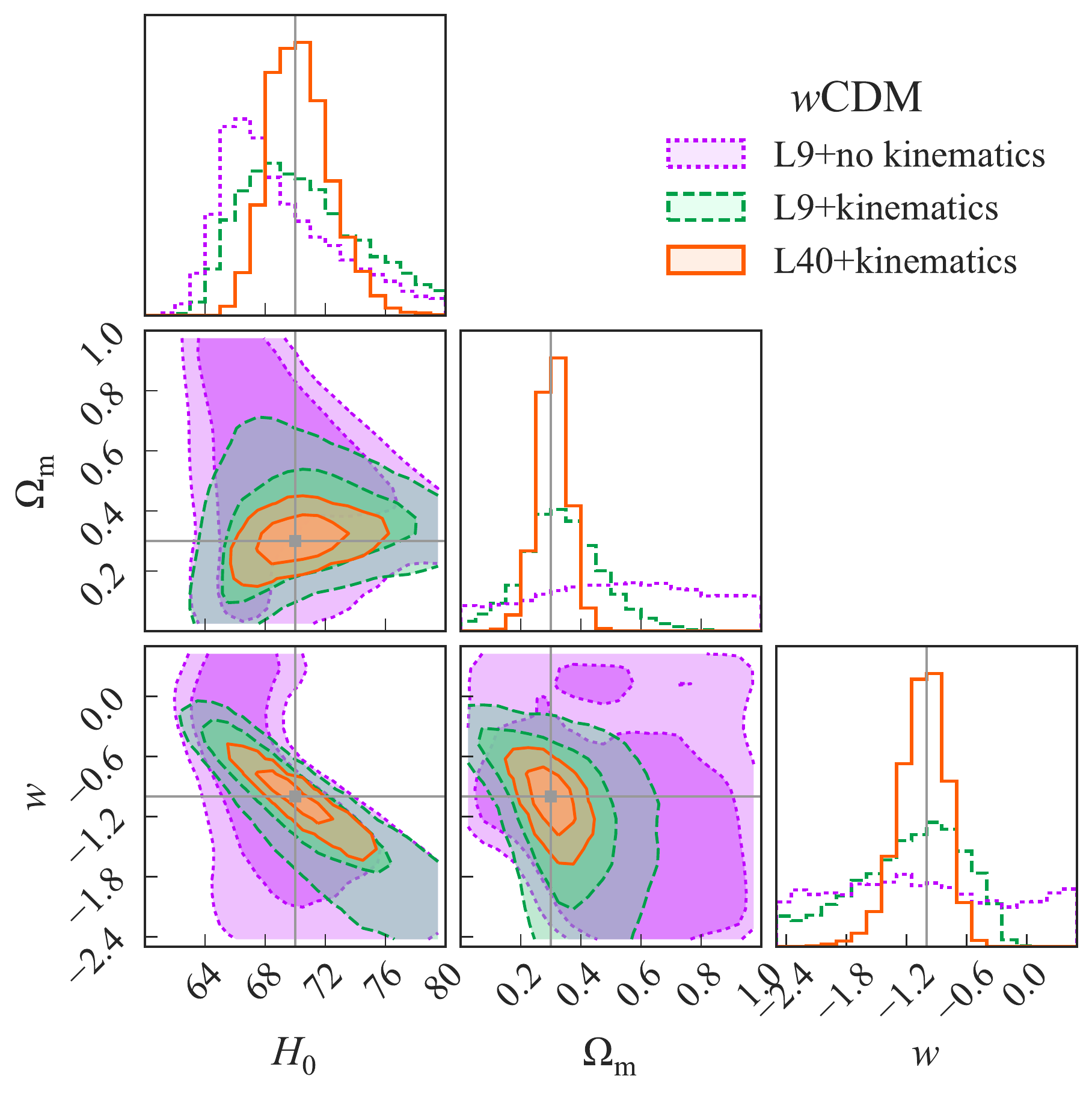} \\
	\includegraphics[width=1\columnwidth]{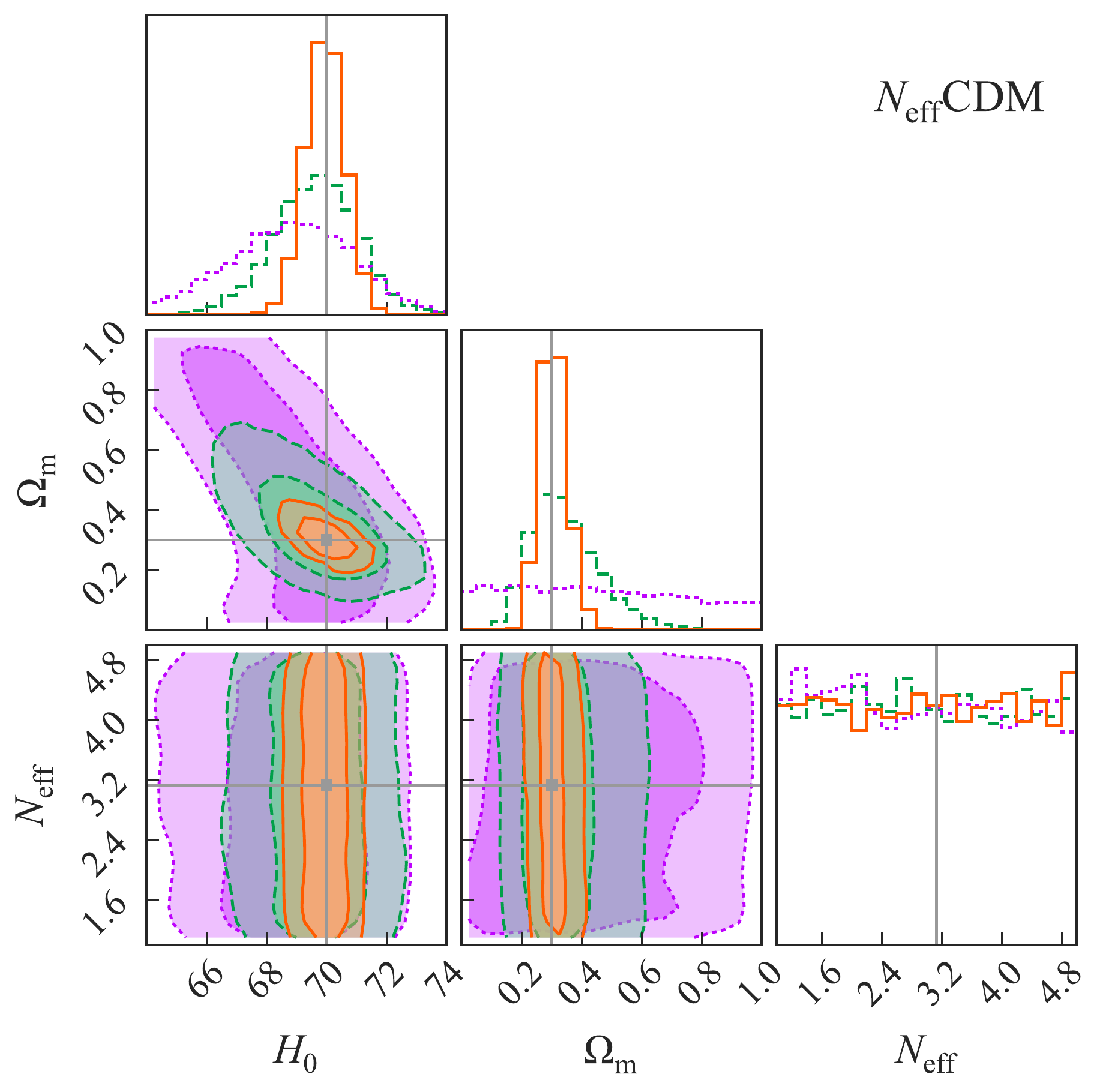}
	\caption{
	Posterior PDF of cosmological parameters obtained from distance measurements for o$\Lambda$CDM (top left), $w$CDM model (top right), and $N_{{\rm eff}}$CDM (bottom) models. The posterior PDF inferred from nine lenses (L9) without kinematics is shown in dotted contours, from nine lenses with kinematics is shown in dashed contours, and from 40 lenses (L40) with kinematics is shown in solid contours. The contours represent 1$\sigma$ and 2$\sigma$ confidence regions. Solid straight lines show the fiducial values. For all cosmological models, adding spatially resolved kinematics lifts degeneracies between the cosmological parameters and puts tighter constraints on them.
	\label{fig:owLCDM_9lens}
	}
\end{figure*}

\begin{figure*}
	\includegraphics[width=2\columnwidth]{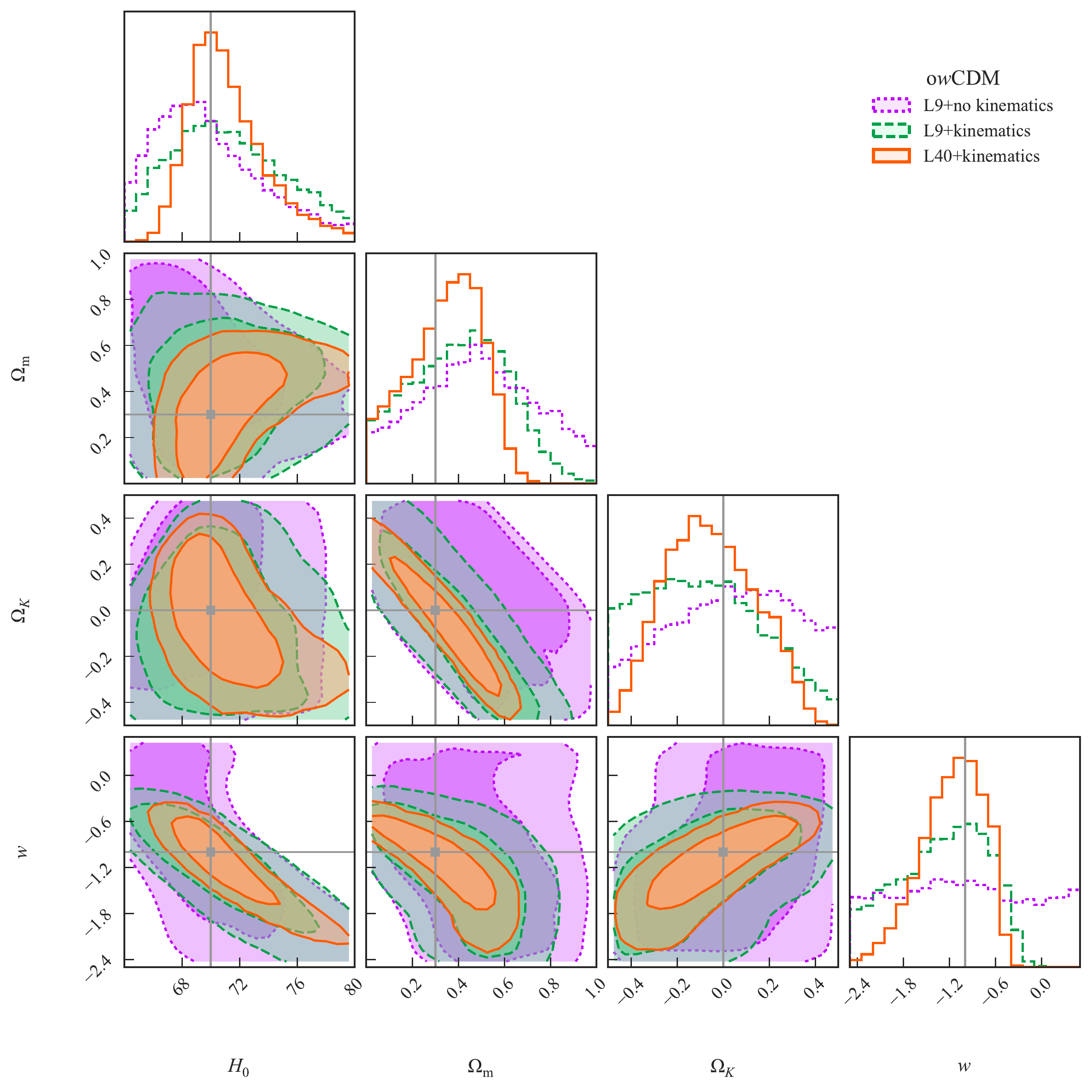}
	\caption{
	Posterior PDF of cosmological parameters obtained from the distance measurements for o$w$CDM model. The posterior PDF inferred from nine lenses (L9) without kinematics is shown in dotted contours, from nine lenses with kinematics is shown in dashed contours, and from 40 lenses (L40) with kinematics is shown in solid contours. The contours represent 1$\sigma$ and 2$\sigma$ confidence regions. Solid straight lines show the fiducial values. This is a further illustration of the role of spatially resolved kinematics in breaking degeneracies between the parameters for a two-parameter extension from the basic $\Lambda$CDM model.
	\label{fig:owLCDM_40lens}
	}
\end{figure*}

\subsubsection{Nine lenses}
The detailed parameter uncertainties for all the cosmological models considered in this paper are tabulated in \autoref{tab:cosmo_params}. For the flat $\Lambda$CDM model, $H_0$ is estimated with 2.0 per cent precision ($69.7 \pm 1.4$ km/s/Mpc) and $\Omega_{{\rm m}}$ is estimated with precision $\sigma(\Omega_{{\rm m}}) = 0.11$.

To measure the improvement over cosmological parameter uncertainties
by using spatially resolved kinematics, we did the same analysis for
nine lenses without using kinematics. In that case, the parameter
uncertainties are $\sigma(H_0) = 3.2$ per cent and
$\sigma(\Omega_{{\rm m}}) = 0.32$. Using
spatially resolved kinematics for nine lenses leads to an improvement
in the precision of $H_0$ from 3.2 to 2.0 per cent. If we adopt the conservative lensing data quality equivalent to $\delta \gamma \sim 0.04$, addition of the spatially resolved stellar kinematics for nine lenses still improves the precision of $H_0$ by 1 per cent from 3.9 to 2.9 per cent. Without any kinematics there is a very strong degeneracy in $\Omega_{\rm m}$, which can be broken by adding the stellar kinematics information (\autoref{fig:LCDM_9lens}).

For the o$\Lambda$CDM model with our ``primary'' fiducial cosmology, the cosmological parameter uncertainties are estimated to be $\sigma(H_0) = 3.3 $ per cent, $\sigma(\Omega_{{\rm m}}) = 0.2$, $\sigma(\Omega_{{\rm K}}) = 0.27$ (\autoref{fig:owLCDM_9lens}). For the flat $w$CDM model, we estimate the cosmological parameters uncertainties to be $\sigma(H_0) = 6.2$ per cent, $\sigma(\Omega_{{\rm m}}) =0.13$, and $\sigma(w) = 0.57$ (\autoref{fig:owLCDM_9lens}). For the $N_{{\rm eff}}$CDM model, the parameter uncertainties are estimated to be $\sigma(H_0)=2.0$ per cent, $\sigma(\Omega_{{\rm m}})=0.11$ and $N_{\rm eff}$ is completely degenerate.

For the o$w$CDM model, where we relax $\Omega_{K}$ and $w$ from the flat $\Lambda$CDM model, we estimate the parameters with uncertainties $\sigma(H_0) = 6.5$ per cent, $\sigma(\Omega_{{\rm m}}) = 0.22$, $\sigma(\Omega_{K}) = 0.28$,  $\sigma(w) = 0.63$. For the $w_a$CDM model, $w_0$ and $w_a$ are estimated with uncertainties $\sigma(w_0)=0.82$ and $\sigma(w_a)=3.5$, respectively.
% giving a dark-energy figure-of-merit (FoM, given by $1/\sigma(w_0)\sigma(w_a)$ for uncorrelated $w_0$ and $w_a$) of $0.31$. 

%{\bf TT: is it true that H0 is more precise in owCDM than in wCDM? I am a bit suprised}

\subsubsection{40 lenses}

For the flat $\Lambda$CDM model, using distance measurement uncertainties from 40 lenses we estimate $H_0$ with 0.92 per cent precision and $\Omega_{{\rm m}}$ with $\sigma(\Omega_{{\rm m}}) = 0.044 $. For the conservative lensing data quality equivalent to $\delta \gamma \sim 0.04$, the sample of 40 lenses constraints $H_0$ with 1.3 per cent uncertainty. The parameter uncertainties are estimated for o$\Lambda$CDM model to be $\sigma(H_0)=1.6$ per cent, $\sigma(\Omega_{{\rm m}})=0.089$, and $\sigma(\Omega_{K}) = 0.12$ and for $w$CDM model to be $\sigma(H_0)=2.9$ per cent, $\sigma(\Omega_{{\rm m}})=0.05$, and $\sigma(w) = 0.22$. For $N_{{\rm eff}}$CDM model, we estimate the parameter uncertainties to be $\sigma(H_0)=0.93$ per cent, $\sigma(\Omega_{{\rm m}})=0.045$. Adding more lens to the sample does not improve the degeneracy in $N_{\rm eff}$ showing time-delay cosmography is insensitive to $N_{\rm eff}$.

For the o$w$CDM model, we estimate the parameter uncertainties to be $\sigma(H_0) = 3.6$ per cent, $\sigma(\Omega_{{\rm m}}) = 0.17$, $\sigma(\Omega_{{\rm K}}) = 0.21$,  $\sigma(w) = 0.42$ (\autoref{fig:owLCDM_40lens}). For the $w_a$CDM model, $w_0$ and $w_a$ are estimated with uncertainties $\sigma(w_0)=0.55$ and $\sigma(w_a)=3.0$, respectively. %, leading to FoM = $0.51$.

\subsection{Joint analysis with \textit{Planck}}
\label{ssec:joint}

We combined the inference on cosmography from strong lensing with \textit{Planck} 2015 data release \citep[][hereafter \textit{Planck}]{Planck15XIII}.\footnote{We used the \textit{Planck} chains designated by ``plikHM\_TT\_lowTEB'' which uses the baseline high-l \textit{Planck} power spectra and low-l temperature and LFI polarization.} To combine the two data sets, we followed the importance sampling method prescribed by \citet{Lewis02} and implemented by \citet{Suyu10, Suy++13}, and \citet{Bonvin17}. We used the bivariate normal distribution fit of the posterior PDF of $D_{{\rm d}}$ and $D_{\Delta t}$ given in Equation \eqref{eq:gaussianfit} to compute the ``importance'' or weight of each point in the \textit{Planck} chain.

\begin{figure*}
	\includegraphics[width=\columnwidth]{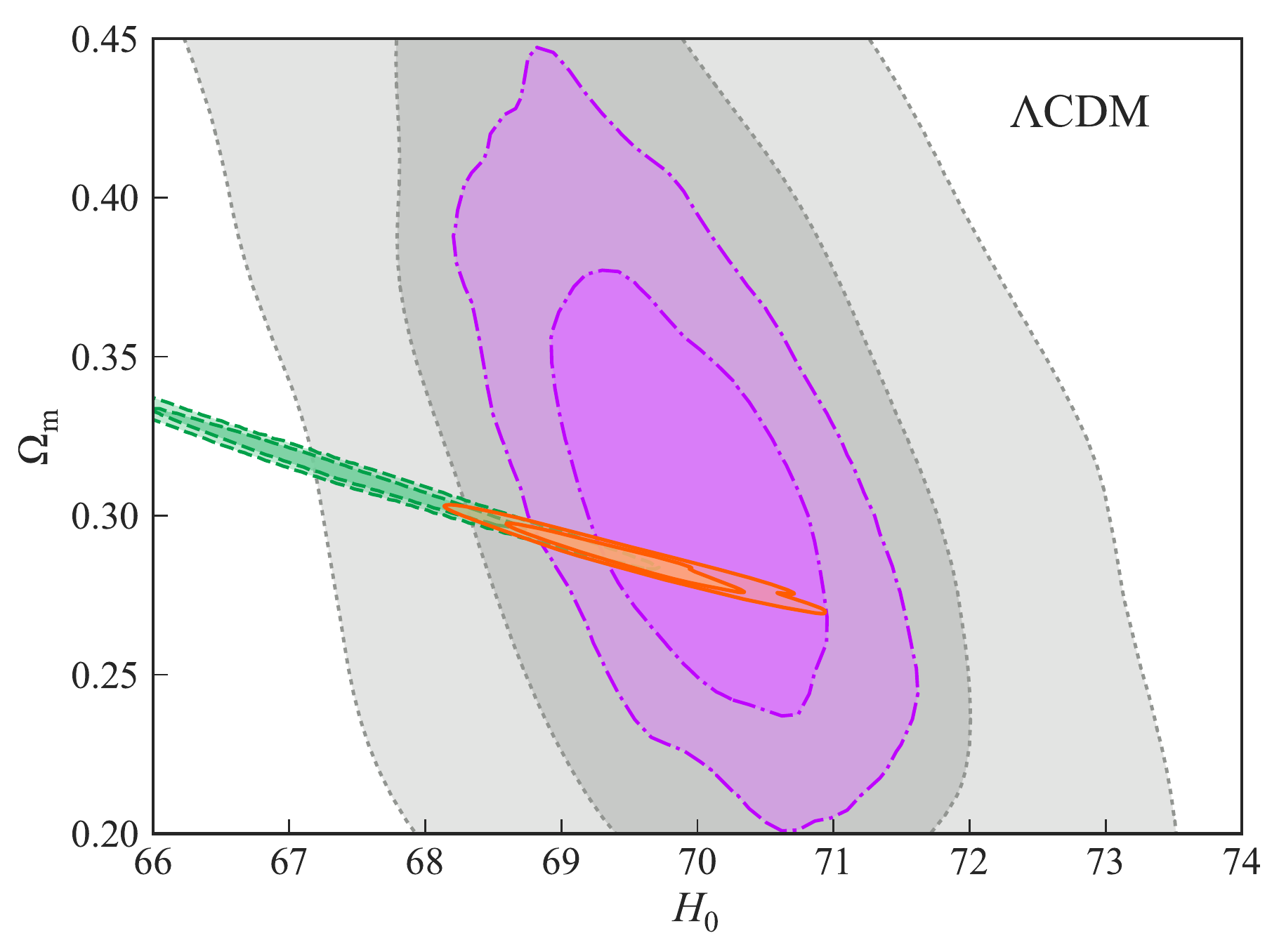}
	\includegraphics[width=\columnwidth]{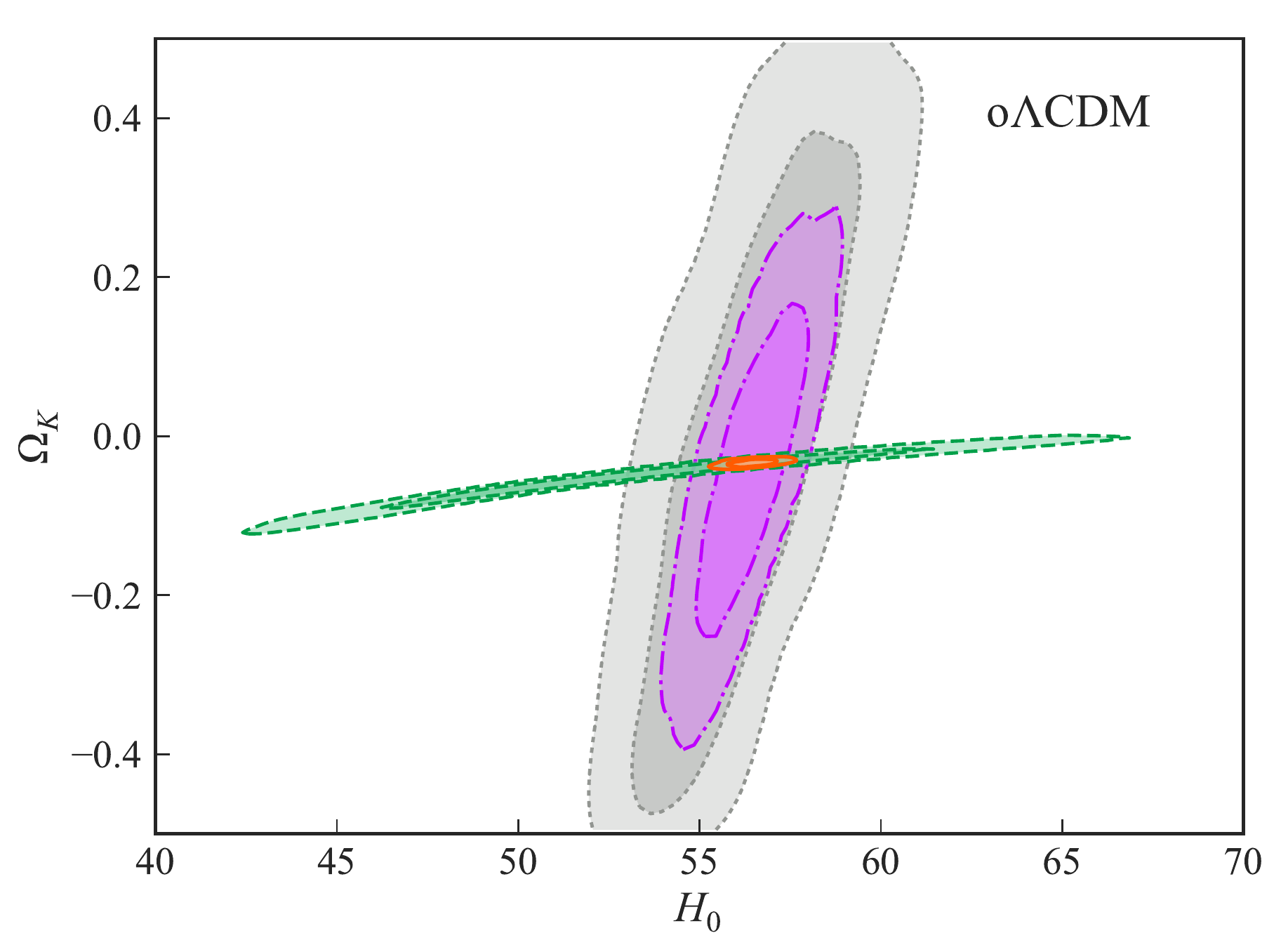} \\
	\includegraphics[width=\columnwidth]{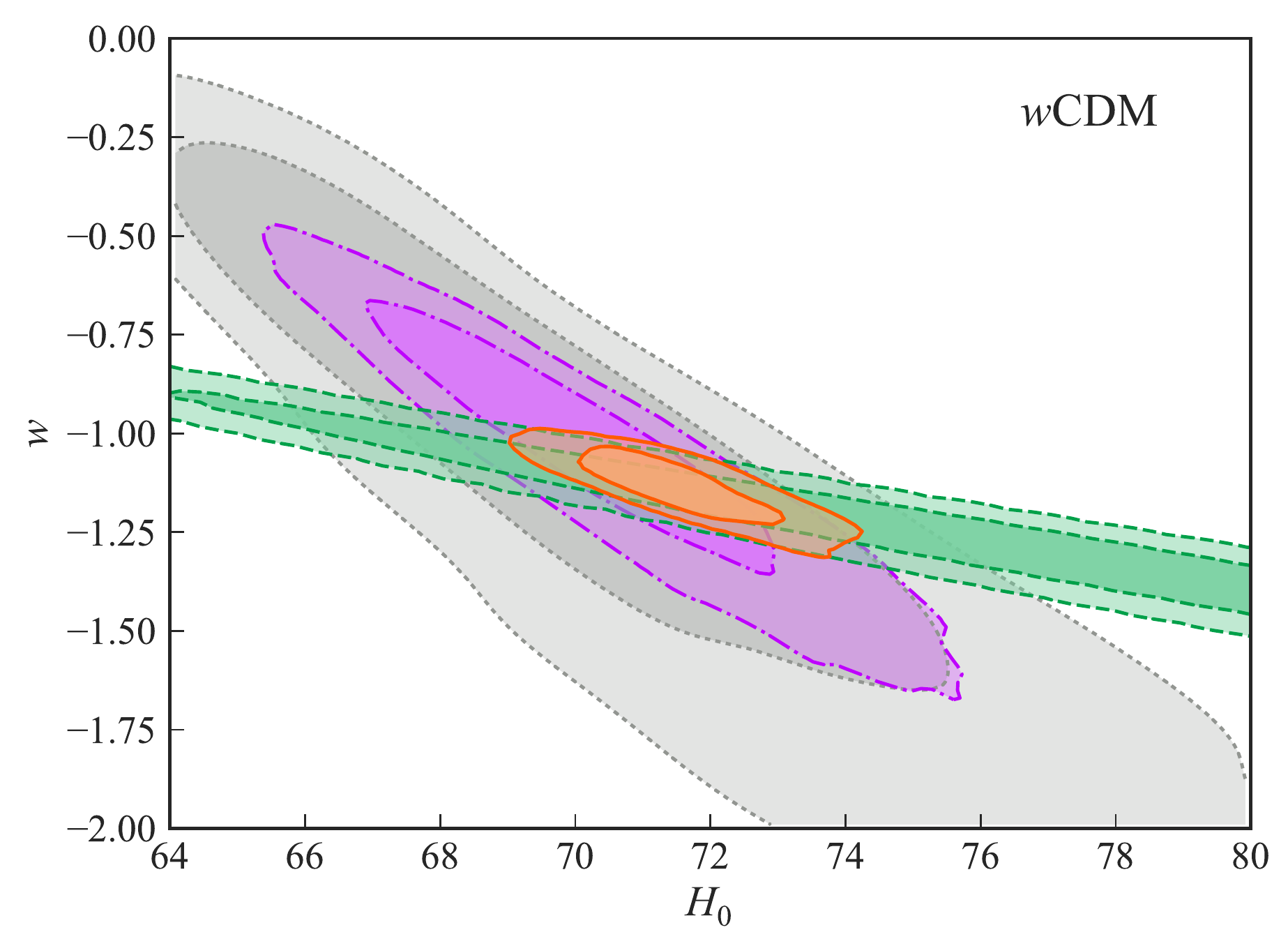}
	\includegraphics[width=\columnwidth]{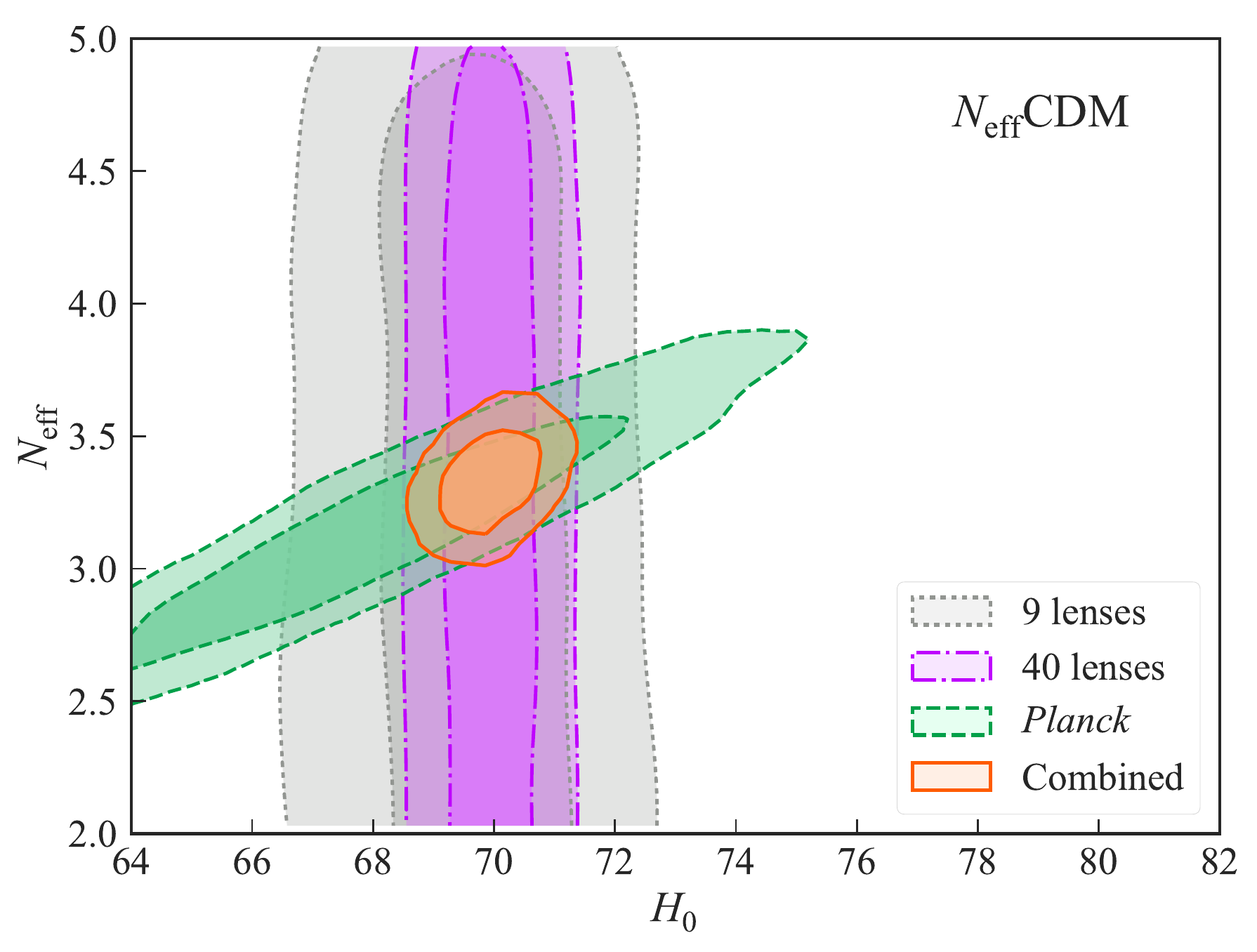}
	
	\caption{
	1$\sigma$ and 2$\sigma$ regions of cosmological parameters obtained from lensing data alone and in combination with \textit{Planck} for $\Lambda$CDM (top left), o$\Lambda$CDM (top right), $w$CDM (bottom left), and $N_{{\rm eff}}$CDM (bottom right) models. The constraints from nine lenses with spatially resolved kinematics are shown with dotted contours, from 40 lenses with spatially resolved kinematics are shown with dash-dotted contours, from \textit{Planck} are shown in dashed contours, and from the combination of \textit{Planck} and 40 lenses are shown in solid contours. In all cases, adding the lensing information to the \textit{Planck} data improves the constraints on the cosmological parameters.
	\label{fig:lens_planck_comb}
	}
\end{figure*}

For many combinations of cosmological model and parameters, the confidence regions from the time-delay cosmography are orthogonal to the ones from the \textit{Planck}. As a result, combining the inferences from the time-delay cosmography with the \textit{Planck} leads to much tighter constraints (\autoref{fig:lens_planck_comb}).

For flat $\Lambda$CDM model, combining \textit{Planck} with nine lenses leads to an 1.1 per cent measurement of $H_0$. For the combination of 40 lenses and \textit{Planck}, the precision of $H_0$ becomes 0.74 per cent (\autoref{tab:cosmo_params}) in the flat $\Lambda$CDM model. 

For o$\Lambda$CDM model, the maximum likelihood regions of the \textit{Planck} and the lensing data with the ``primary'' fiducial cosmology are too far apart to implement the importance sampling method. Therefore, we used the \textit{Planck} values, $H_0 = 56.5$ km/s/Mpc, $\Omega_{{\rm m}} = 0.441$, $\Omega_{\Lambda} = 0.592$, $\Omega_{K} = -0.033$ as the fiducial cosmology to generate the lensing likelihood to combine with the \textit{Planck} likelihood. This combination gives $\sigma(H_0)=1.8$ per cent, $\sigma({\Omega_{{\rm m}}}) = 0.017$, and $\sigma(\Omega_{K}) = 0.0048$ for nine lenses and $\sigma(H_0)=0.81$ per cent, $\sigma({\Omega_{{\rm m}}}) = 0.01$, and $\sigma(\Omega_{K}) = 0.0031$ for 40 lenses.

For the $w$CDM model, the precision of $w$ is estimated to be $\sigma(w) = 0.081$ and $\sigma(w) = 0.060$ for combination of \textit{Planck} with nine and 40 lenses, respectively. For the $N_{{\rm eff}}$CDM model, we constrain the number of relativistic species with $\sigma(N_{{\rm eff}}) = 0.16$ and $\sigma(N_{{\rm eff}}) = 0.13$ by combining \textit{Planck} with nine and 40 lenses, respectively.

\begin{figure}
	\includegraphics[width=\columnwidth]{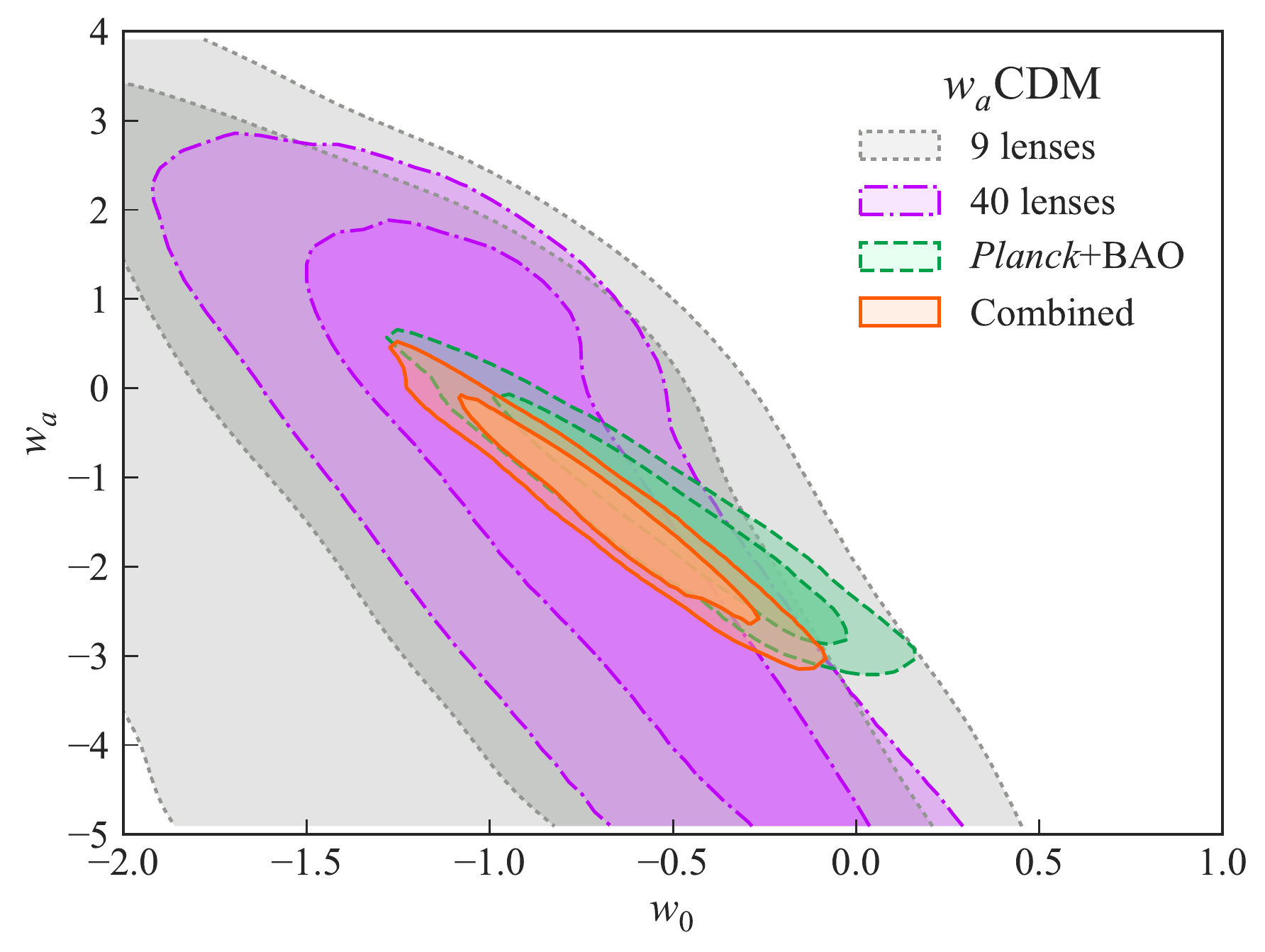}
	\caption{
	1$\sigma$ and 2$\sigma$ confidence regions of the dark energy equation of state parameters obtained from lensing data alone and in combination with \textit{Planck} for $w_a$CDM model. The constraints from nine lenses with spatially resolved kinematics are shown with dotted contours, from 40 lenses with spatially resolved kinematics are shown with dash-dotted contours, from \textit{Planck} are shown in dashed contours, and from the combination of \textit{Planck} and 40 lenses are shown in solid contours. Adding lensing measurements to the \textit{Planck}+BAO data improves the dark energy FoM by 56 per cent.
	\label{fig:lens_planck_comb_wwa}
	}
\end{figure}

We did not combine \textit{Planck} with the lensing likelihoods for  o$\Lambda$CDM model as the \textit{Planck} collaboration did not publicly release the parameter chains for this model. For $w_a$CDM model, we combined the lensing information with \textit{Planck}+BAO constraints. From the joint analysis, we estimate the parameter uncertainties to be $\sigma(w_0) = 0.29$ and $\sigma(w_a)=0.86$ giving dark-energy figure of merit (FoM; given by the inverse of the area enclosed by the 1$\sigma$ confidence contour in the $w_0-w_a$ plane) 0.85 for nine lenses, and $\sigma(w_0) = 0.27$ and $\sigma(w_a)=0.82$ giving an FoM = 1.11 for 40 lenses (\autoref{fig:lens_planck_comb_wwa}).

\section{Discussion and Comparison with Previous Work}
\label{sect:discussion}

We explored how incorporating spatially resolved kinematics of the
deflector in addition to the lensing and time-delay data improves the
precision of the inferred cosmological parameters. We showed that the
addition of the spatially resolved kinematics to the lensing and
time-delay data helps break the mass-anisotropy degeneracy and leads
to improved precision in the determination of the angular diameter
distance of the deflector (from $\sim20$ to $\sim 10$ per cent). We found
that the time-delay distances can be simultaneously measured with
$\sim6$ per cent accuracy, which is comparable to the 6 per cent measurement of the
time-delay distance for the lens RXJ1131-1231 \citep{Suy++13} and the
7.6 per cent measurement for the lens HE0435-1223 \citep{Wong17}. These
precision margins are achievable by current and future IFSs, e.g.
OSIRIS on Keck with laser guide star AO or space-based instruments,
e.g. NIRSPEC on \textit{JWST}. Future telescopes like TMT or E-ELT would
improve these precisions further. \citet{Jee16} assume 5 per cent precision
on both angular diameter and time-delay distances; however we found
that 5 per cent precision on angular diameter distance measurement is
probably beyond reach with current or imminent technology.

We confirmed the result by \citet{Linder11} and \citet{Jee16} that combining lensing information with CMB and other external data sets can greatly improve the constraints on the cosmological parameters. \citet{Linder11} finds that by combining time-delay distance measurements with 5 per cent uncertainty from 150 hypothetical strong lens systems with the CMB and supernova constraints, dark energy FoM can be improved by almost a factor of 5. \citet{Jee16} find that combining angular diameter and time-delay distance measurements with 5 per cent uncertainties on both from 55 lenses with \textit{Planck}+BAO+JLA constraints improves the precision on $w_a$ and the dark energy figure of merit by approximately a factor of 2 for the time-varying dark energy model. In our study, combining angular diameter and time-delay distance measurements with $\sim$10 per cent and $\sim$6 per cent uncertainties, respectively, from 40 lenses with \textit{Planck}+BAO data improves the \textit{Planck}+BAO constraint on $w_a$ by 13 per cent and the dark energy FoM by 56 per cent, consistent with previous results after taking into account the differences.

\section{Limitation of this present work}\label{sect:limitation}

In order to model a large number of lenses in a computationally
efficient manner we adopted many simplifying assumptions. First,
we used a collection of conjugate points to replace the detailed
modelling of the lensed quasar host galaxy. Secondly, we assumed
spherical symmetry to speed up the calculations. By requiring the
uncertainty on the effective mass density profile slope to be equal to
0.02, the precision obtained by full-blown models, this procedure
ensures that we get realistic uncertainties on distances. We know
from full pixel-based simulations that such precision can be attained
by modelling images obtained with reasonable exposure time using
current and future technology \citep{Meng++15}. A similar study is
required to estimate the exposure times required to carry out the
spectroscopic observations (Meng et al. 2017, in preparation).

%We have adopted a simplified 
%Simulating a strong gravitational lens to its truest nature would be computationally costly for two reasons: (1) complex and diverse nature of the deflector mass distribution, and (2) necessity to model the intricate images of the lensed quasar and its host galaxy. Several models for galaxy mass distribution have been proposed including those that generalize beyond the simple mass profiles \citep[e.g.][]{Zhao96, Munoz01}, and those that incorporate ellipticity in the model \citep[e.g.][]{Golse02, Meneghetti++03}. Modeling an elliptical galaxy to its most generalized form greatly increase the computational load by incorporating extra dimensions to the parameter space and by introducing complex functions to compute which in some cases can not be analytically treated. We assumed the spherical cases of the Jaffe and the NFW profiles which are not only empirically motivated, but also computationally feasible to produce realistic mock data within an acceptable computing time. In addition, a full-blown pixel-based analysis of the lensed image data can take up a significant amount of CPU time to complete just for one system, let alone for the 40 systems considered in this study. We used a grid of conjugate points with 10 mas separation to mimic the extended lensed images and calibrated the resolution to produce realistic estimate of the model parameters while minimizing the CPU time. 

We assumed a 3-5 per cent uncertainty for the external convergence as it can be indirectly estimated by comparing the statistics of galaxy number counts along the line of sight with simulated light cones from the Millennium Simulation. This approach has the caveat of being dependent on the assumed cosmology and thus possibly biasing the final cosmological inferences \citep{Rus++16}. Moreover, there can be $\sim$25 per cent deviation in the inferred $\kappa_{\rm ext}$ between the \textit{Planck} cosmology and the Millennium Simulation. This would leave some residual systematics to be accounted for when analysing real-life lenses. With the most pessimistic approach of 25 per cent variation between median $\kappa_{\rm ext}$ inferred from ray-tracing, this would mean that a median value of $\kappa_{\rm ext}=0.04$ would impart a 1 per cent systematic uncertainty on $H_0$. However, $\kappa_{\rm ext}$ can be shown to depend primarily on $\Omega_{\rm m}\sigma_{8}$ where $\sigma_8$ is the root-mean-square fluctuation of the mass density, while other contributions are sub-dominant \citep[Equation C4 in][]{Rus++16}. This means that one can perform a complete cosmographic inference, where also $\Omega_{\rm m}$ and $\sigma_{8}$ are varied when importance-sampling from the CMB chains. Whereas the product $\Omega_{\rm m}\sigma_{8}$ (hence the reconstructed median $\kappa_{\rm ext}$) can vary appreciably between ``different'' CMB experiments (with different setups, or different multipole coverage, or beam characterization), its possible variation is smaller within the same CMB experiment, which means that the median $\kappa_{\rm ext}$ inferred will vary by less than the most pessimistic estimate (25 to 1 per cent on $H_0$). Hence, regardless of whether $\Omega_{\rm m}\sigma_{8}$ are varied or kept at a fiducial value when considering $\kappa_{\rm ext}$, time-delays are still a robust way of probing departures from flat-$\Lambda$CDM as inferred from within a particular CMB probe, without particular worries from the $\kappa_{\rm ext}$ reconstruction, with sub-percent accuracy. There are, in fact, other factors affecting the accuracy of $\kappa_{\rm ext}$ reconstruction, such as the choice of weighting scheme in terms of distances and masses, or the importance of a multi-plane approach. However, when dealing with real-life lenses, these have been (and are being) discussed at length for each individual system while still at blinded-inference stage. Different lenses have required different evaluations of $\kappa_{\rm ext}$, but after unblinding they have all given consistent $H_0$ results, which suggests that this side of reconstruction systematics is currently under control. Part of the reason may be that the width of the $\kappa_{\rm ext}$ PDF is not negligible with respect to the median, so any systematics affecting the shift in $\kappa_{\rm ext}$ are still comprised within 1$\sigma$ from the ``true'' value.

Finally, we emphasize that our study takes into account systematic
uncertainties only in an indirect manner. Possible sources of known
systematics can be contamination from the bright quasar images to the the
host galaxy flux or the deflector spectra, unaccounted line-of-sight contribution to the projected mass etc. We assumed
that these sources of known systematics can be accounted by our chosen
error budget for different mock data and model parameters, e.g.
5-10 per cent uncertainty on the velocity dispersion and 3-5 per cent uncertainty on
the external convergence, which are realistic error budgets for these
quantities from the state of the art measurements. It would be useful
to carry out a systematic investigation of strategies that may be
required to limit any potential bias arising from these systematic
uncertainties to be well below the statistical errors.

\section{Summary}
\label{sect:summary}

Strong lenses with measured time delays can be used as probes to constrain cosmological parameters through the measurement of the angular diameter distance to the deflector and the time-delay distance. However, spatially resolved kinematics is essential to measure the angular diameter distance to the deflector and it also helps break the mass-anisotropy degeneracy. In this paper, we used a realistic model of a deflector galaxy to create mock lensing and kinematic data and estimated the achievable precisions of the cosmological parameters. The main conclusions of this study are as follows

\begin{enumerate}
	\item The angular diameter distance to the deflector can be measured to approximately 10 per cent uncertainty by including spatially resolved kinematics from current ground-based IFS with laser guide star AO, e.g. OSIRIS on Keck, or with space-based instruments, e.g. NIRSPEC on \textit{JWST}, to the imaging data of the lensed quasar and the time-delay measurement. The time-delay distance can be simultaneously measured to $\sim$6 per cent uncertainty.
	
	\item Using spatially resolved kinematics improves the
          precision on angular diameter distance per system from $\sim$20 to $\sim$10 per cent over using integrated kinematic data.

  \item $H_0$ can be measured to 2.0 per cent precision using lensing and spatially resolved kinematics from nine lenses and to sub-percent precision (0.9 per cent) from 40 lenses.

	\item The uncertainty on $H_0$ improves from 3.2 to 2.0 per cent by adding the spatially resolved kinematics to the lensing and time-delay data for nine strong lens systems.
	
	\item Combining \textit{Planck} with lensing and spatially
          resolved kinematics data can break degeneracies between the
          cosmological parameters and greatly improve the constraints
          on them. Especially, for the time-dependent dark energy
          parameter model, the dark energy FoM is improved
          by 56 per cent by combining a sample of 40 lenses with
          measured time delays and kinematics with \textit{Planck}+BAO constraints.
	\end{enumerate}
	
        This is a very interesting time for time-delay cosmography as
        several wide-field and deep-sky surveys such as the Dark
        Energy Survey, \textit{Euclid}, the Wide Field Infrared Survey
        Telescope (WFIRST), the Large Synoptic Survey Telescope
        (LSST), should allow for the discovery and follow-up of tens
        to hundreds of multiply imaged quasars \citep{O+M10}.  The
        launch of NIRSPEC on \textit{JWST}, scheduled for 2018, and the
        constantly improving ground-based instruments with laser guide
        star AO (e.g. OSIRIS on Keck) make it possible to have
        high-quality spatially resolved kinematics for these lens
        systems. In turn, this can lead to the measurement of the
        Hubble parameter to sub-per-cent precision. Combining the
        distance measurements from strong lens systems to other
        cosmological probes, i.e. CMB, BAO, and Type Ia supernova,
        would help tightly constraint the cosmological parameters
        leading to a deeper understanding of dark energy, dark matter,
        and other fundamental properties of our Universe.

\section*{Acknowledgment}

We thank Simon Birrer, Inh Jee, Eiichiro Komatsu, Philip J. Marshall, Sherry H. Suyu, and Xin Wang for many insightful conversations.  We also thank the anonymous referee whose comments helped us to improve this work. AS and TT
acknowledge support by the Packard Foundation through a Packard
Research Fellowship and by the National Science Foundation through
grant AST-1450141. This work used computational and storage services associated with the Hoffman2 Shared Cluster provided by UCLA Institute for Digital Research and Education's Research Technology Group. This research made use of Astropy, a community-developed core \textsc{Python} package for Astronomy \citep{astropy}, and the \textsc{corner.py} code at \url{https://github.com/dfm/corner.py} \citep{corner}.

%% Bibliography
\bibliographystyle{mnras}
\bibliography{myrefs}

\begin{thebibliography}{}
\makeatletter
\relax
\def\mn@urlcharsother{\let\do\@makeother \do\$\do\&\do\#\do\^\do\_\do\%\do\~}
\def\mn@doi{\begingroup\mn@urlcharsother \@ifnextchar [ {\mn@doi@}
  {\mn@doi@[]}}
\def\mn@doi@[#1]#2{\def\@tempa{#1}\ifx\@tempa\@empty \href
  {http://dx.doi.org/#2} {doi:#2}\else \href {http://dx.doi.org/#2} {#1}\fi
  \endgroup}
\def\mn@eprint#1#2{\mn@eprint@#1:#2::\@nil}
\def\mn@eprint@arXiv#1{\href {http://arxiv.org/abs/#1} {{\tt arXiv:#1}}}
\def\mn@eprint@dblp#1{\href {http://dblp.uni-trier.de/rec/bibtex/#1.xml}
  {dblp:#1}}
\def\mn@eprint@#1:#2:#3:#4\@nil{\def\@tempa {#1}\def\@tempb {#2}\def\@tempc
  {#3}\ifx \@tempc \@empty \let \@tempc \@tempb \let \@tempb \@tempa \fi \ifx
  \@tempb \@empty \def\@tempb {arXiv}\fi \@ifundefined
  {mn@eprint@\@tempb}{\@tempb:\@tempc}{\expandafter \expandafter \csname
  mn@eprint@\@tempb\endcsname \expandafter{\@tempc}}}

\bibitem[\protect\citeauthoryear{{Agnello}, {Auger}  \& {Evans}}{{Agnello}
  et~al.}{2013}]{AAE13}
{Agnello} A.,  {Auger} M.~W.,   {Evans} N.~W.,  2013, \mn@doi [\mnras]
  {10.1093/mnrasl/sls020}, \href
  {http://adsabs.harvard.edu/abs/2013MNRAS.429L..35A} {429, L35}

\bibitem[\protect\citeauthoryear{{Agnello}, {Evans}  \& {Romanowsky}}{{Agnello}
  et~al.}{2014a}]{AER14}
{Agnello} A.,  {Evans} N.~W.,   {Romanowsky} A.~J.,  2014a, \mn@doi [\mnras]
  {10.1093/mnras/stu959}, \href
  {http://adsabs.harvard.edu/abs/2014MNRAS.442.3284A} {442, 3284}

\bibitem[\protect\citeauthoryear{{Agnello}, {Evans}, {Romanowsky}  \&
  {Brodie}}{{Agnello} et~al.}{2014b}]{Agn++14}
{Agnello} A.,  {Evans} N.~W.,  {Romanowsky} A.~J.,   {Brodie} J.~P.,  2014b,
  \mn@doi [\mnras] {10.1093/mnras/stu960}, \href
  {http://adsabs.harvard.edu/abs/2014MNRAS.442.3299A} {442, 3299}

\bibitem[\protect\citeauthoryear{Akeret, Seehars, Amara, Refregier  \&
  Csillaghy}{Akeret et~al.}{2013}]{Ake++12}
Akeret J.,  Seehars S.,  Amara A.,  Refregier A.,   Csillaghy A.,  2013,
  \mn@doi [Astronomy and Computing] {10.1016/j.ascom.2013.06.003}, 2, 27

\bibitem[\protect\citeauthoryear{{Alam} et~al.,}{{Alam}
  et~al.}{2017}]{Shadab16}
{Alam} S.,  et~al., 2017, \mn@doi [\mnras] {10.1093/mnras/stx721}, \href
  {http://adsabs.harvard.edu/abs/2017MNRAS.470.2617A} {470, 2617}

\bibitem[\protect\citeauthoryear{{Astropy Collaboration} et~al.,}{{Astropy
  Collaboration} et~al.}{2013}]{astropy}
{Astropy Collaboration} et~al., 2013, \mn@doi [\aap]
  {10.1051/0004-6361/201322068}, \href
  {http://adsabs.harvard.edu/abs/2013A%26A...558A..33A} {558, A33}

\bibitem[\protect\citeauthoryear{{Auger}, {Treu}, {Bolton}, {Gavazzi},
  {Koopmans}, {Marshall}, {Moustakas}  \& {Burles}}{{Auger}
  et~al.}{2010}]{Auger10}
{Auger} M.~W.,  {Treu} T.,  {Bolton} A.~S.,  {Gavazzi} R.,  {Koopmans}
  L.~V.~E.,  {Marshall} P.~J.,  {Moustakas} L.~A.,   {Burles} S.,  2010,
  \mn@doi [\apj] {10.1088/0004-637X/724/1/511}, \href
  {http://adsabs.harvard.edu/abs/2010ApJ...724..511A} {724, 511}

\bibitem[\protect\citeauthoryear{{Barnab{\`e}}, {Czoske}, {Koopmans}, {Treu}
  \& {Bolton}}{{Barnab{\`e}} et~al.}{2011}]{Bar++11}
{Barnab{\`e}} M.,  {Czoske} O.,  {Koopmans} L.~V.~E.,  {Treu} T.,   {Bolton}
  A.~S.,  2011, \mn@doi [\mnras] {10.1111/j.1365-2966.2011.18842.x}, \href
  {http://adsabs.harvard.edu/abs/2011MNRAS.415.2215B} {415, 2215}

\bibitem[\protect\citeauthoryear{{Bartelmann}}{{Bartelmann}}{1996}]{Bart96}
{Bartelmann} M.,  1996, \aap, \href
  {http://adsabs.harvard.edu/abs/1996A%26A...313..697B} {313, 697}

\bibitem[\protect\citeauthoryear{{Bartelmann} \& {Meneghetti}}{{Bartelmann} \&
  {Meneghetti}}{2004}]{B+M04}
{Bartelmann} M.,  {Meneghetti} M.,  2004, \mn@doi [\aap]
  {10.1051/0004-6361:20035763}, \href
  {http://adsabs.harvard.edu/abs/2004A%26A...418..413B} {418, 413}

\bibitem[\protect\citeauthoryear{{Bernal}, {Verde}  \& {Riess}}{{Bernal}
  et~al.}{2016}]{BVR16}
{Bernal} J.~L.,  {Verde} L.,   {Riess} A.~G.,  2016, \mn@doi [\jcap]
  {10.1088/1475-7516/2016/10/019}, \href
  {http://adsabs.harvard.edu/abs/2016JCAP...10..019B} {10, 019}

\bibitem[\protect\citeauthoryear{{Birrer}, {Amara}  \& {Refregier}}{{Birrer}
  et~al.}{2016}]{BAR16}
{Birrer} S.,  {Amara} A.,   {Refregier} A.,  2016, \mn@doi [\jcap]
  {10.1088/1475-7516/2016/08/020}, \href
  {http://adsabs.harvard.edu/abs/2016JCAP...08..020B} {8, 020}

\bibitem[\protect\citeauthoryear{{Bonvin} et~al.,}{{Bonvin}
  et~al.}{2017}]{Bonvin17}
{Bonvin} V.,  et~al., 2017, \mn@doi [\mnras] {10.1093/mnras/stw3006}, \href
  {http://adsabs.harvard.edu/abs/2017MNRAS.465.4914B} {465, 4914}

\bibitem[\protect\citeauthoryear{{Brewer}, {Marshall}, {Auger}, {Treu},
  {Dutton}  \& {Barnab{\`e}}}{{Brewer} et~al.}{2014}]{Bre++14}
{Brewer} B.~J.,  {Marshall} P.~J.,  {Auger} M.~W.,  {Treu} T.,  {Dutton} A.~A.,
    {Barnab{\`e}} M.,  2014, \mn@doi [\mnras] {10.1093/mnras/stt2026}, \href
  {http://adsabs.harvard.edu/abs/2014MNRAS.437.1950B} {437, 1950}

\bibitem[\protect\citeauthoryear{{Chevallier} \& {Polarski}}{{Chevallier} \&
  {Polarski}}{2001}]{Chevallier01}
{Chevallier} M.,  {Polarski} D.,  2001, \mn@doi [International Journal of
  Modern Physics D] {10.1142/S0218271801000822}, \href
  {http://adsabs.harvard.edu/abs/2001IJMPD..10..213C} {10, 213}

\bibitem[\protect\citeauthoryear{{Collett} et~al.,}{{Collett}
  et~al.}{2013}]{Col++13}
{Collett} T.~E.,  et~al., 2013, \mn@doi [\mnras] {10.1093/mnras/stt504}, \href
  {http://adsabs.harvard.edu/abs/2013MNRAS.432..679C} {432, 679}

\bibitem[\protect\citeauthoryear{{Courteau} et~al.,}{{Courteau}
  et~al.}{2014}]{Cou++14}
{Courteau} S.,  et~al., 2014, \mn@doi [Reviews of Modern Physics]
  {10.1103/RevModPhys.86.47}, \href
  {http://adsabs.harvard.edu/abs/2014RvMP...86...47C} {86, 47}

\bibitem[\protect\citeauthoryear{{Dutton} \& {Treu}}{{Dutton} \&
  {Treu}}{2014}]{D+T14}
{Dutton} A.~A.,  {Treu} T.,  2014, \mn@doi [\mnras] {10.1093/mnras/stt2489},
  \href {http://adsabs.harvard.edu/abs/2014MNRAS.438.3594D} {438, 3594}

\bibitem[\protect\citeauthoryear{{Eisenstein} et~al.,}{{Eisenstein}
  et~al.}{2005}]{Eisenstein05}
{Eisenstein} D.~J.,  et~al., 2005, \mn@doi [\apj] {10.1086/466512}, \href
  {http://adsabs.harvard.edu/abs/2005ApJ...633..560E} {633, 560}

\bibitem[\protect\citeauthoryear{{Falco}, {Gorenstein}  \& {Shapiro}}{{Falco}
  et~al.}{1985}]{FGS85}
{Falco} E.~E.,  {Gorenstein} M.~V.,   {Shapiro} I.~I.,  1985, \mn@doi [\apjl]
  {10.1086/184422}, \href {http://adsabs.harvard.edu/abs/1985ApJ...289L...1F}
  {289, L1}

\bibitem[\protect\citeauthoryear{Foreman-Mackey}{Foreman-Mackey}{2016}]{corner}
Foreman-Mackey D.,  2016, \mn@doi [The Journal of Open Source Software]
  {10.21105/joss.00024}, 24

\bibitem[\protect\citeauthoryear{{Foreman-Mackey}, {Hogg}, {Lang}  \&
  {Goodman}}{{Foreman-Mackey} et~al.}{2013}]{Emcee13}
{Foreman-Mackey} D.,  {Hogg} D.~W.,  {Lang} D.,   {Goodman} J.,  2013, \mn@doi
  [\pasp] {10.1086/670067}, \href
  {http://adsabs.harvard.edu/abs/2013PASP..125..306F} {125, 306}

\bibitem[\protect\citeauthoryear{{Gavazzi}, {Treu}, {Rhodes}, {Koopmans},
  {Bolton}, {Burles}, {Massey}  \& {Moustakas}}{{Gavazzi}
  et~al.}{2007}]{Gav++07}
{Gavazzi} R.,  {Treu} T.,  {Rhodes} J.~D.,  {Koopmans} L.~V.~E.,  {Bolton}
  A.~S.,  {Burles} S.,  {Massey} R.~J.,   {Moustakas} L.~A.,  2007, \mn@doi
  [\apj] {10.1086/519237}, \href
  {http://adsabs.harvard.edu/abs/2007ApJ...667..176G} {667, 176}

\bibitem[\protect\citeauthoryear{{Gavazzi}, {Treu}, {Koopmans}, {Bolton},
  {Moustakas}, {Burles}  \& {Marshall}}{{Gavazzi} et~al.}{2008}]{Gav++08}
{Gavazzi} R.,  {Treu} T.,  {Koopmans} L.~V.~E.,  {Bolton} A.~S.,  {Moustakas}
  L.~A.,  {Burles} S.,   {Marshall} P.~J.,  2008, \mn@doi [\apj]
  {10.1086/529541}, \href {http://adsabs.harvard.edu/abs/2008ApJ...677.1046G}
  {677, 1046}

\bibitem[\protect\citeauthoryear{Goodman \& Weare}{Goodman \&
  Weare}{2010}]{G+W10}
Goodman J.,  Weare J.,  2010, \mn@doi [Communications in Applied Mathematics
  and Computational Science] {10.2140/camcos.2010.5.65}, 5, 65

\bibitem[\protect\citeauthoryear{{Greene} et~al.,}{{Greene}
  et~al.}{2013}]{Gre++13}
{Greene} Z.~S.,  et~al., 2013, \mn@doi [\apj] {10.1088/0004-637X/768/1/39},
  \href {http://adsabs.harvard.edu/abs/2013ApJ...768...39G} {768, 39}

\bibitem[\protect\citeauthoryear{{Grillo}, {Lombardi}  \& {Bertin}}{{Grillo}
  et~al.}{2008}]{GLB08}
{Grillo} C.,  {Lombardi} M.,   {Bertin} G.,  2008, \mn@doi [\aap]
  {10.1051/0004-6361:20077534}, \href
  {http://adsabs.harvard.edu/abs/2008A%26A...477..397G} {477, 397}

\bibitem[\protect\citeauthoryear{{Hilbert}, {Hartlap}, {White}  \&
  {Schneider}}{{Hilbert} et~al.}{2009}]{Hil++09}
{Hilbert} S.,  {Hartlap} J.,  {White} S.~D.~M.,   {Schneider} P.,  2009,
  \mn@doi [\aap] {10.1051/0004-6361/200811054}, \href
  {http://adsabs.harvard.edu/abs/2009A%26A...499...31H} {499, 31}

\bibitem[\protect\citeauthoryear{{Jaffe}}{{Jaffe}}{1983}]{Jaffe83}
{Jaffe} W.,  1983, \mn@doi [\mnras] {10.1093/mnras/202.4.995}, \href
  {http://adsabs.harvard.edu/abs/1983MNRAS.202..995J} {202, 995}

\bibitem[\protect\citeauthoryear{{Jee}, {Komatsu}  \& {Suyu}}{{Jee}
  et~al.}{2015}]{Jee15}
{Jee} I.,  {Komatsu} E.,   {Suyu} S.~H.,  2015, \mn@doi [\jcap]
  {10.1088/1475-7516/2015/11/033}, \href
  {http://adsabs.harvard.edu/abs/2015JCAP...11..033J} {11, 033}

\bibitem[\protect\citeauthoryear{{Jee}, {Komatsu}, {Suyu}  \& {Huterer}}{{Jee}
  et~al.}{2016}]{Jee16}
{Jee} I.,  {Komatsu} E.,  {Suyu} S.~H.,   {Huterer} D.,  2016, \mn@doi [\jcap]
  {10.1088/1475-7516/2016/04/031}, \href
  {http://adsabs.harvard.edu/abs/2016JCAP...04..031J} {4, 031}

\bibitem[\protect\citeauthoryear{{Kennedy} \& {Eberhart}}{{Kennedy} \&
  {Eberhart}}{1995}]{Kennedy95}
{Kennedy} J.,  {Eberhart} R.,  1995, in Proceedings of
  {ICNN}{\textquotesingle}95 - International Conference on Neural Networks.
  {IEEE}, \mn@doi{10.1109/icnn.1995.488968}, \url
  {https://doi.org/10.1109/icnn.1995.488968}

\bibitem[\protect\citeauthoryear{{Koopmans}, {Treu}, {Fassnacht}, {Blandford}
  \& {Surpi}}{{Koopmans} et~al.}{2003}]{Koo++03}
{Koopmans} L.~V.~E.,  {Treu} T.,  {Fassnacht} C.~D.,  {Blandford} R.~D.,
  {Surpi} G.,  2003, \mn@doi [\apj] {10.1086/379226}, \href
  {http://adsabs.harvard.edu/abs/2003ApJ...599...70K} {599, 70}

\bibitem[\protect\citeauthoryear{{Koopmans}, {Treu}, {Bolton}, {Burles}  \&
  {Moustakas}}{{Koopmans} et~al.}{2006}]{Koo++06}
{Koopmans} L.~V.~E.,  {Treu} T.,  {Bolton} A.~S.,  {Burles} S.,   {Moustakas}
  L.~A.,  2006, \mn@doi [\apj] {10.1086/505696}, \href
  {http://adsabs.harvard.edu/abs/2006ApJ...649..599K} {649, 599}

\bibitem[\protect\citeauthoryear{{Koopmans} et~al.,}{{Koopmans}
  et~al.}{2009}]{Koo++09}
{Koopmans} L.~V.~E.,  et~al., 2009, \mn@doi [\apjl]
  {10.1088/0004-637X/703/1/L51}, \href
  {http://adsabs.harvard.edu/abs/2009ApJ...703L..51K} {703, L51}

\bibitem[\protect\citeauthoryear{{Larkin} et~al.,}{{Larkin}
  et~al.}{2006}]{Lar++06}
{Larkin} J.,  et~al., 2006, \mn@doi [\nar] {10.1016/j.newar.2006.02.005}, \href
  {http://adsabs.harvard.edu/abs/2006NewAR..50..362L} {50, 362}

\bibitem[\protect\citeauthoryear{{Lewis} \& {Bridle}}{{Lewis} \&
  {Bridle}}{2002}]{Lewis02}
{Lewis} A.,  {Bridle} S.,  2002, \mn@doi [\prd] {10.1103/PhysRevD.66.103511},
  \href {http://adsabs.harvard.edu/abs/2002PhRvD..66j3511L} {66, 103511}

\bibitem[\protect\citeauthoryear{{Linder}}{{Linder}}{2003}]{Linder03}
{Linder} E.~V.,  2003, \mn@doi [Physical Review Letters]
  {10.1103/PhysRevLett.90.091301}, \href
  {http://adsabs.harvard.edu/abs/2003PhRvL..90i1301L} {90, 091301}

\bibitem[\protect\citeauthoryear{{Linder}}{{Linder}}{2011}]{Linder11}
{Linder} E.~V.,  2011, \mn@doi [\prd] {10.1103/PhysRevD.84.123529}, \href
  {http://adsabs.harvard.edu/abs/2011PhRvD..84l3529L} {84, 123529}

\bibitem[\protect\citeauthoryear{{Mamon} \& {{\L}okas}}{{Mamon} \&
  {{\L}okas}}{2005}]{Mamon05}
{Mamon} G.~A.,  {{\L}okas} E.~L.,  2005, \mn@doi [\mnras]
  {10.1111/j.1365-2966.2005.09400.x}, \href
  {http://adsabs.harvard.edu/abs/2005MNRAS.363..705M} {363, 705}

\bibitem[\protect\citeauthoryear{{Meneghetti}, {Bartelmann}  \&
  {Moscardini}}{{Meneghetti} et~al.}{2003}]{MBM03}
{Meneghetti} M.,  {Bartelmann} M.,   {Moscardini} L.,  2003, \mn@doi [\mnras]
  {10.1046/j.1365-8711.2003.06276.x}, \href
  {http://adsabs.harvard.edu/abs/2003MNRAS.340..105M} {340, 105}

\bibitem[\protect\citeauthoryear{{Meng}, {Treu}, {Agnello}, {Auger}, {Liao}  \&
  {Marshall}}{{Meng} et~al.}{2015}]{Meng++15}
{Meng} X.-L.,  {Treu} T.,  {Agnello} A.,  {Auger} M.~W.,  {Liao} K.,
  {Marshall} P.~J.,  2015, \mn@doi [\jcap] {10.1088/1475-7516/2015/09/059},
  \href {http://adsabs.harvard.edu/abs/2015JCAP...09..059M} {9, 059}

\bibitem[\protect\citeauthoryear{{Merritt}}{{Merritt}}{1985a}]{Merritt85a}
{Merritt} D.,  1985a, \mn@doi [\aj] {10.1086/113810}, \href
  {http://adsabs.harvard.edu/abs/1985AJ.....90.1027M} {90, 1027}

\bibitem[\protect\citeauthoryear{{Merritt}}{{Merritt}}{1985b}]{Merritt85b}
{Merritt} D.,  1985b, \mn@doi [\mnras] {10.1093/mnras/214.1.25P}, \href
  {http://adsabs.harvard.edu/abs/1985MNRAS.214P..25M} {214, 25P}

\bibitem[\protect\citeauthoryear{{Navarro}, {Frenk}  \& {White}}{{Navarro}
  et~al.}{1996}]{Navarro96}
{Navarro} J.~F.,  {Frenk} C.~S.,   {White} S.~D.~M.,  1996, \mn@doi [\apj]
  {10.1086/177173}, \href {http://adsabs.harvard.edu/abs/1996ApJ...462..563N}
  {462, 563}

\bibitem[\protect\citeauthoryear{{Navarro}, {Frenk}  \& {White}}{{Navarro}
  et~al.}{1997}]{NFW97}
{Navarro} J.~F.,  {Frenk} C.~S.,   {White} S.~D.~M.,  1997, \mn@doi [\apj]
  {10.1086/304888}, \href {http://adsabs.harvard.edu/abs/1997ApJ...490..493N}
  {490, 493}

\bibitem[\protect\citeauthoryear{{Oguri} \& {Marshall}}{{Oguri} \&
  {Marshall}}{2010}]{O+M10}
{Oguri} M.,  {Marshall} P.~J.,  2010, \mn@doi [\mnras]
  {10.1111/j.1365-2966.2010.16639.x}, \href
  {http://adsabs.harvard.edu/abs/2010MNRAS.405.2579O} {405, 2579}

\bibitem[\protect\citeauthoryear{{Osipkov}}{{Osipkov}}{1979}]{Osipkov79}
{Osipkov} L.~P.,  1979, Pisma v Astronomicheskii Zhurnal, \href
  {http://adsabs.harvard.edu/abs/1979PAZh....5...77O} {5, 77}

\bibitem[\protect\citeauthoryear{{Paraficz} \& {Hjorth}}{{Paraficz} \&
  {Hjorth}}{2009}]{Paraficz10}
{Paraficz} D.,  {Hjorth} J.,  2009, \mn@doi [\aap]
  {10.1051/0004-6361/200913307}, \href
  {http://adsabs.harvard.edu/abs/2009A%26A...507L..49P} {507, L49}

\bibitem[\protect\citeauthoryear{{Perlmutter} et~al.,}{{Perlmutter}
  et~al.}{1999}]{Perlmutter98}
{Perlmutter} S.,  et~al., 1999, \mn@doi [\apj] {10.1086/307221}, \href
  {http://adsabs.harvard.edu/abs/1999ApJ...517..565P} {517, 565}

\bibitem[\protect\citeauthoryear{{Planck Collaboration} et~al.,}{{Planck
  Collaboration} et~al.}{2016}]{Planck15XIII}
{Planck Collaboration} et~al., 2016, \mn@doi [\aap]
  {10.1051/0004-6361/201525830}, \href
  {http://adsabs.harvard.edu/abs/2016A%26A...594A..13P} {594, A13}

\bibitem[\protect\citeauthoryear{{Refsdal}}{{Refsdal}}{1964}]{Refsdal64}
{Refsdal} S.,  1964, \mn@doi [\mnras] {10.1093/mnras/128.4.307}, \href
  {http://adsabs.harvard.edu/abs/1964MNRAS.128..307R} {128, 307}

\bibitem[\protect\citeauthoryear{{Riess} et~al.,}{{Riess}
  et~al.}{1998}]{Riess98}
{Riess} A.~G.,  et~al., 1998, \mn@doi [\aj] {10.1086/300499}, \href
  {http://adsabs.harvard.edu/abs/1998AJ....116.1009R} {116, 1009}

\bibitem[\protect\citeauthoryear{{Riess} et~al.,}{{Riess}
  et~al.}{2016}]{Riess16}
{Riess} A.~G.,  et~al., 2016, \mn@doi [\apj] {10.3847/0004-637X/826/1/56},
  \href {http://adsabs.harvard.edu/abs/2016ApJ...826...56R} {826, 56}

\bibitem[\protect\citeauthoryear{{Rusu} et~al.,}{{Rusu} et~al.}{2017}]{Rus++16}
{Rusu} C.~E.,  et~al., 2017, \mn@doi [\mnras] {10.1093/mnras/stx285}, \href
  {http://adsabs.harvard.edu/abs/2017MNRAS.467.4220R} {467, 4220}

\bibitem[\protect\citeauthoryear{{Schneider} \& {Sluse}}{{Schneider} \&
  {Sluse}}{2013}]{S+S13}
{Schneider} P.,  {Sluse} D.,  2013, \mn@doi [\aap]
  {10.1051/0004-6361/201321882}, \href
  {http://adsabs.harvard.edu/abs/2013A%26A...559A..37S} {559, A37}

\bibitem[\protect\citeauthoryear{{Schneider} \& {Sluse}}{{Schneider} \&
  {Sluse}}{2014}]{SPT}
{Schneider} P.,  {Sluse} D.,  2014, \mn@doi [\aap]
  {10.1051/0004-6361/201322106}, \href
  {http://adsabs.harvard.edu/abs/2014A%26A...564A.103S} {564, A103}

\bibitem[\protect\citeauthoryear{{Schneider}, {Kochanek}  \&
  {Wambsganss}}{{Schneider} et~al.}{2006}]{Schneider06}
{Schneider} P.,  {Kochanek} C.~S.,   {Wambsganss} J.,  2006, Gravitational
  Lensing: Strong, Weak and Micro.
 Saas-Fee Advanced Courses Vol. 33, Springer

\bibitem[\protect\citeauthoryear{{Sluse} et~al.,}{{Sluse}
  et~al.}{2017}]{Slu++16}
{Sluse} D.,  et~al., 2017, \mn@doi [\mnras] {10.1093/mnras/stx1484}, \href
  {http://adsabs.harvard.edu/abs/2017MNRAS.470.4838S} {470, 4838}

\bibitem[\protect\citeauthoryear{{Suyu}}{{Suyu}}{2012}]{Suy12}
{Suyu} S.~H.,  2012, \mn@doi [\mnras] {10.1111/j.1365-2966.2012.21661.x}, \href
  {http://adsabs.harvard.edu/abs/2012MNRAS.426..868S} {426, 868}

\bibitem[\protect\citeauthoryear{{Suyu}, {Marshall}, {Auger}, {Hilbert},
  {Blandford}, {Koopmans}, {Fassnacht}  \& {Treu}}{{Suyu}
  et~al.}{2010}]{Suyu10}
{Suyu} S.~H.,  {Marshall} P.~J.,  {Auger} M.~W.,  {Hilbert} S.,  {Blandford}
  R.~D.,  {Koopmans} L.~V.~E.,  {Fassnacht} C.~D.,   {Treu} T.,  2010, \mn@doi
  [\apj] {10.1088/0004-637X/711/1/201}, \href
  {http://adsabs.harvard.edu/abs/2010ApJ...711..201S} {711, 201}

\bibitem[\protect\citeauthoryear{{Suyu} et~al.,}{{Suyu} et~al.}{2012}]{Suy++12}
{Suyu} S.~H.,  et~al., 2012, preprint, \href
  {http://adsabs.harvard.edu/abs/2012arXiv1202.4459S} {} (\mn@eprint {arXiv}
  {1202.4459})

\bibitem[\protect\citeauthoryear{{Suyu} et~al.,}{{Suyu} et~al.}{2013}]{Suy++13}
{Suyu} S.~H.,  et~al., 2013, \mn@doi [\apj] {10.1088/0004-637X/766/2/70}, \href
  {http://adsabs.harvard.edu/abs/2013ApJ...766...70S} {766, 70}

\bibitem[\protect\citeauthoryear{{Suyu} et~al.,}{{Suyu} et~al.}{2014}]{Suy++14}
{Suyu} S.~H.,  et~al., 2014, \mn@doi [\apjl] {10.1088/2041-8205/788/2/L35},
  \href {http://adsabs.harvard.edu/abs/2014ApJ...788L..35S} {788, L35}

\bibitem[\protect\citeauthoryear{{Treu} \& {Koopmans}}{{Treu} \&
  {Koopmans}}{2002a}]{T+K02b}
{Treu} T.,  {Koopmans} L.~V.~E.,  2002a, \mn@doi [\mnras]
  {10.1046/j.1365-8711.2002.06107.x}, \href
  {http://adsabs.harvard.edu/cgi-bin/nph-bib_query?bibcode=2002MNRAS.337L...6T&db_key=AST}
  {337, L6}

\bibitem[\protect\citeauthoryear{{Treu} \& {Koopmans}}{{Treu} \&
  {Koopmans}}{2002b}]{Treu02}
{Treu} T.,  {Koopmans} L.~V.~E.,  2002b, \mn@doi [\apj] {10.1086/341216}, \href
  {http://adsabs.harvard.edu/abs/2002ApJ...575...87T} {575, 87}

\bibitem[\protect\citeauthoryear{{Treu} \& {Koopmans}}{{Treu} \&
  {Koopmans}}{2004}]{T+K04}
{Treu} T.,  {Koopmans} L.~V.~E.,  2004, \mn@doi [\apj] {10.1086/422245}, \href
  {http://adsabs.harvard.edu/abs/2004ApJ...611..739T} {611, 739}

\bibitem[\protect\citeauthoryear{{Treu} \& {Marshall}}{{Treu} \&
  {Marshall}}{2016}]{Treu16}
{Treu} T.,  {Marshall} P.~J.,  2016, \mn@doi [\aapr]
  {10.1007/s00159-016-0096-8}, \href
  {http://adsabs.harvard.edu/abs/2016A%26ARv..24...11T} {24, 11}

\bibitem[\protect\citeauthoryear{{Weinberg}, {Mortonson}, {Eisenstein},
  {Hirata}, {Riess}  \& {Rozo}}{{Weinberg} et~al.}{2013}]{Wei++13}
{Weinberg} D.~H.,  {Mortonson} M.~J.,  {Eisenstein} D.~J.,  {Hirata} C.,
  {Riess} A.~G.,   {Rozo} E.,  2013, \mn@doi [\physrep]
  {10.1016/j.physrep.2013.05.001}, \href
  {http://adsabs.harvard.edu/abs/2013PhR...530...87W} {530, 87}

\bibitem[\protect\citeauthoryear{{Wong} et~al.,}{{Wong} et~al.}{2017}]{Wong17}
{Wong} K.~C.,  et~al., 2017, \mn@doi [\mnras] {10.1093/mnras/stw3077}, \href
  {http://adsabs.harvard.edu/abs/2017MNRAS.465.4895W} {465, 4895}

\bibitem[\protect\citeauthoryear{{Wright} et~al.,}{{Wright}
  et~al.}{2016}]{Wri++16}
{Wright} S.~A.,  et~al., 2016, in Society of Photo-Optical Instrumentation
  Engineers (SPIE) Conference Series. p. 990905 (\mn@eprint {arXiv}
  {1608.01696}), \mn@doi{10.1117/12.2233182}

\bibitem[\protect\citeauthoryear{{Wucknitz}}{{Wucknitz}}{2002}]{Wuc02}
{Wucknitz} O.,  2002, \mn@doi [\mnras] {10.1046/j.1365-8711.2002.05426.x},
  \href {http://adsabs.harvard.edu/abs/2002MNRAS.332..951W} {332, 951}

\bibitem[\protect\citeauthoryear{{Xu}, {Sluse}, {Schneider}, {Springel},
  {Vogelsberger}, {Nelson}  \& {Hernquist}}{{Xu} et~al.}{2016}]{XuEtal2016}
{Xu} D.,  {Sluse} D.,  {Schneider} P.,  {Springel} V.,  {Vogelsberger} M.,
  {Nelson} D.,   {Hernquist} L.,  2016, \mn@doi [\mnras]
  {10.1093/mnras/stv2708}, \href
  {http://adsabs.harvard.edu/abs/2016MNRAS.456..739X} {456, 739}

\bibitem[\protect\citeauthoryear{{van der Marel}}{{van der
  Marel}}{1994}]{vdM94}
{van der Marel} R.~P.,  1994, \mn@doi [\mnras] {10.1093/mnras/270.2.271}, \href
  {http://adsabs.harvard.edu/abs/1994MNRAS.270..271V} {270, 271}

\makeatother
\end{thebibliography}

\bsp	% typesetting comment 

\label{lastpage}
\end{document}